\documentclass[twocolumn,showkeys,superscriptaddress,preprintnumbers,amsmath,amssymb,showpacs,prx,longbibliography]{revtex4-2}
\usepackage{graphicx}
\usepackage{dcolumn}
\usepackage{bm}
\usepackage{amsmath}
\usepackage{amssymb}
\usepackage{color}
\usepackage[dvipsnames]{xcolor}
\usepackage{hyperref}
\hypersetup{
    colorlinks=true,
    linkcolor=blue,
    filecolor=black,      
    urlcolor=blue,
    citecolor=blue,
}
 \usepackage{natbib}
\usepackage{braket}
\usepackage{soul}
\usepackage{CJK}
\usepackage{relsize}
\usepackage{printlen}
\usepackage{pifont}
\def\X#1{
        \raisebox{.9pt}{\textcircled{\raisebox{-.9pt}{#1}}}
}

\definecolor{testcolor}{rgb}{1    0.5    0}

\begin{document}

\title{Ultra-stable shear jammed granular materials}

\begin{CJK*}{UTF8}{gbsn}
\author{Yiqiu Zhao (赵逸秋)}
\email{yiqiuzhao@ust.hk}
\affiliation{Department of Physics \& Center for Non-linear and Complex Systems, Duke University, Durham, NC, 27708}
\affiliation{Department of Physics, The Hong Kong University of Science and Technology, Hong Kong SAR, China}

\author{Yuchen Zhao (赵雨辰)}
\affiliation{Department of Physics \& Center for Non-linear and Complex Systems, Duke University, Durham, NC, 27708}
\affiliation{School of Mechanical and Aerospace Engineering, Nanyang Technological University, 639798, Singapore}

\author{Dong Wang (王东)}
\affiliation{Department of Physics \& Center for Non-linear and Complex Systems, Duke University, Durham, NC, 27708}
\affiliation{Department of Mechanical Engineering \& Materials Science, Yale University, New Haven, CT, 06520}

\author{Hu~Zheng~(郑虎)}
\affiliation{Department of Physics \& Center for Non-linear and Complex Systems, Duke University, Durham, NC, 27708}
\affiliation{Department of Geotechnical Engineering, College of Civil Engineering, Tongji University, Shanghai, China}

\author{Bulbul Chakraborty}
\affiliation{Martin Fisher School of Physics, Brandeis University, Waltham, MA, 02454}

\author{Joshua~E.~S.~Socolar}
\email{socolar@duke.edu}
\affiliation{Department of Physics \& Center for Non-linear and Complex Systems, Duke University, Durham, NC, 27708}

\date{\today}


\begin{abstract}
Dry granular materials such as sand, gravel, pills, or agricultural grains, can become rigid when compressed or sheared. Under isotropic compression, the material will reach a certain {\it jamming} density and then resist further compression. {\em Shear jamming} occurs when resistance to shear emerges in a system at a density lower than the jamming density. Although shear-jamming is prevalent in frictional granular materials, their stability properties are not well described by standard elasticity theory and thus call for experimental characterization. We report on experimental observations of changes in the mechanical properties of a shear-jammed granular material subjected to small-amplitude, quasi-static cyclic shear. We study a layer of plastic discs confined to a shear cell, using photoelasticimetry to measure all inter-particle vector forces. For sufficiently small cyclic shear amplitudes and large enough initial shear, the material evolves to an unexpected ``ultra-stable'' state in which all the particle positions and inter-particle contact forces remain unchanged after each complete shear cycle for thousands of cycles. The stress response of these states to small imposed shear is nearly elastic, in contrast to the original shear jammed state. 
\end{abstract}


\maketitle


\section{Introduction}



Granular materials are collections of athermal particles that interact with each other only by contact forces~\cite{Jeager1996_rmp,deGennes1999_rmp}. These materials are ubiquitous in nature and are important components of many industrial products and processes. Under externally imposed stress, a set of grains that flows like a liquid can jam into a solid packing by forming a rigid, disordered contact network~\cite{Liu1998_nature,OHern2003_pre,Majmudar2007_prl,Liu2010_arcmp,Bi2011_nat,behringer2018_rpp}. Jamming is a non-equilibrium process, and the properties of the jammed packing depend on the driving protocol~\cite{Bi2011_nat,dagois2012_prl,Bertrand2016_pre,Baity2017_jps}.

Shear-induced jamming has attracted much attention recently due to its appearance in a large variety of particulate systems~\cite{Bi2011_nat,zhao2019_prl,Kumar2016_gm,urbani2017_prl,Jin2018_sa,otsuki2020_pre,Jine2021_pnas,Xiong2019_gm} and its direct relevance in controlling the discontinuous shear thickening of dense suspension~\cite{mari2014_jor,wyart2014_prl,brown2014_RPP,han2018_prf,blanco2019_pnas,morris2020_arfm}. Shear-jammed granular materials are known to be fragile~\cite{Cates1998_prl}
in contrast to jammed structures induced by compression; their packing structures are highly unstable to changes in the boundary stresses~\cite{Bi2011_nat,zhao2019_prl,seto2019_gm}. While the origin of rigidity of shear-jammed systems has been discussed in recent works~\cite{Bi2011_nat,sarkar2013_prl,sarkar2016_pre,dong2018_prl,Baity2017_jps}, the stability of shear-jammed packings against finite (not infinitesimal) external perturbations remains poorly understood. 

In this work, we examine the stability of shear-jammed packings by monitoring the system evolution under additional strain-controlled shear cycles. We observe a remarkable effect: under certain conditions, sustained cyclic shear leads to a state in which a force network emerges that persists without change over thousands of additional cycles.  The effect is dramatically illustrated in strobe movies of the evolution of the particle positions and the force network.  For some preparations, the movie shows a dense force network in an originally shear jammed packing fade to a completely stress free state, while for others, the network becomes less dense but locks into a steady state, one that 
turns out to be stiffer 
than the original state. (See supplementals videos.)  
In the steady state, under a complete shear cycle, all the particles return to the same position and all the contact forces return to the same state.  
We will call the jammed states in this type of limit cycle ``ultra-stable'' to distinguish them from other jammed states that relax plastically under applied shear cycles. 

We report here on the elastic and yielding properties of these  ``ultra-stable'' states.
To our knowledge, there has been no previous experimental observation of such states
We note that related phenomena have been observed in recent numerical studies of gravitationally stabilized packings \cite{Royer2015_pnas} and packings above the isotropic jamming density~\cite{Otsuki2021_epje}, and also in cyclic shear experiments on micron-sized spheres that interact through electrostatic dipole-dipole interactions~\cite{keim2014_prl,Galloway2022_natphy,keim2021_arxiv}. 
These studies, however, do not speak to origin or stability of the shear-jammed states of interest here. 

\begin{figure}[t]
    \centering
    \includegraphics[width = 0.99\columnwidth]{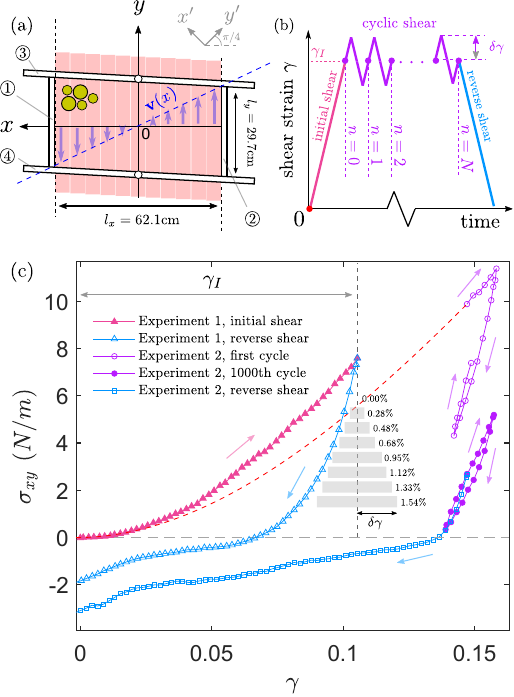}
    \caption{(a) A top view schematic of the multi-slot shear cell. (b) The driving strain protocal. An initial large forward shear is followed by multiple periods of small amplitude cyclic shear, and then by large amplitude reverse shear. The shear rate is always in the quasistatic regime. (c) Shear stress $\sigma_{xy}$ defined in Eq.~(\ref{eq:stress_tensor}) as functions of shear strain $\gamma$ for two example experiments. Experiment 1 contains an initial forward shear $\gamma_{\rm I} = 0.105$ followed directly by an reverse shear.  Experiment 2 contains an initial shear (sketched by the red dashed curve) up to $\gamma_{\rm I} = 0.147$ followed by a series of cyclic shear with $\delta\gamma = 0.95\%$ and then an reverse shear. For experiment 2 only the responses in the first and the 1000th cycle are plotted. The gray boxes display all the cyclic strain amplitudes used in the cyclic shear tests.}
    \label{fig:procedure}
\end{figure}

Our experimental granular system is a monolayer of photoelastic discs set in a special shearing device that is capable of imposing homogeneous internal shear
strains, shown schematically in Fig.~\ref{fig:procedure}(a). We measure all the inter-particle contact forces using photoelasticimetry~\cite{majmudar2005_nature,Daniels2017_rsi,zadeh2019_gm}. Beginning with an unjammed packing, we apply an initial volume conserving shear $\gamma_{\rm I}$ to create a shear jammed state.
We then examine its response to small-amplitude cyclic shear $\delta\gamma$ (See Fig.~\ref{fig:procedure}(b)). The yielding behavior of the shear-jammed and ultra-stable states are examined by reversing the original shear, as indicated in Fig.~\ref{fig:procedure}(b).

We prepare hundreds of shear-jammed packings with different $\gamma_{\rm I}$  beginning from different unjammed configurations and use $\delta\gamma \ll \gamma_{\rm I}$. One may expect a threshold $\delta\gamma$ below which a shear-jammed system behaves elastically. However, such a threshold is usually negligibly small for real-world granular materials~\cite{andreotti_forterre_pouliquen_2013}. 
Under finite strain, a packing typically becomes unstable in the sense that particles rearrange from cycle to cycle, even when boundary stress versus strain curves appear similar. (See, for example, Ref.~\cite{ren2013_thesis}).

Our study shows that, although shear-jammed states are not ultra-stable in general, ultra-stable states may appear after 
{\color{black} a series of}
quasistatic shear cycles is applied to a shear-jammed packing. For a given cyclic strain amplitude $\delta\gamma$, ultra-stable states appear only for  $\gamma_{\rm I}$ larger than a threshold value. For $\gamma_{\rm I}$ below the threshold, the original shear-jammed state becomes unjammed. The ultra-stable shear-jammed packings, on the other hand, behave like anisotropic elastic solids for small shear strains and undergo a sharp yielding transition when strained beyond the cyclic strain amplitude.

This paper is organized as follows.  Section~\ref{sec_methods} describes our experimental apparatus and protocol.  Section~\ref{sec_results} presents our results in three categories: (A) the conditions required for the formation of an ultra-stable state; (B) the different elastic characteristics of the ultra-stable states and the original states; and (C) the yielding transition under large-strain reverse shear. Section~\ref{sec_discussion} contains a summary of our major findings and remarks on their significance.

\section{Materials and experimental protocol}\label{sec_methods}

Our model granular system consists of a bidisperse collection of 1040 photoelastic discs with diameters $d_{\rm b}=15.9\,$mm and $d_{\rm s}=12.7\,$mm and thickness $h=6.8\,$mm. The number ratio of large to small discs is $1/3$, and we keep the system's packing fraction fixed at $\phi=0.816$ throughout all experiments. This value is close to, but less than the isotropic jamming packing fraction $\phi_{\rm J}\approx0.835$~\cite{Ren2013_prl}. The static friction coefficients are  $\mu = 0.87\pm 0.03$ between the particles, $\mu_{\rm base}=0.25\pm0.05$ between a particle and the base, and  $\mu_{\rm wall}= 0.70\pm 0.02$ between a particle and boundary wall. The particles are cut using a water jet from a polyurethane sheet (Precision Urethane \& Machine, Inc.). The bulk modulus of the material is $5.88\,$MPa. Under static diametric loading, the normal contact force law is roughly Hertzian, but some degree of hysteresis is observed under cyclic loading. Details on contact force law calibration are given in Appendix.~\ref{app_deflection}.

We use the multi-slat simple shear apparatus developed by Ren et al.~\cite{Ren2013_prl} to impose a {\it uniform} shear strain field. Our setup avoids the formation of a shear band and the associated density heterogeneity before the jamming onset~\cite{Ren2013_prl}. A schematic top view of the apparatus is shown in Fig.~\ref{fig:procedure}(a). The shear cell contains four aluminum walls (white rectangles) as confining boundaries and a bottom formed by 50 parallel acrylic slats (light purple rectangles, shown as only 11).
Each slat, as well as wall \X1 and wall \X2, are constrained to move only along the $y$ direction, while the two other walls are constrained to rotate with pivots at $(0,l_y/2)$ and $(0,-l_y/2)$.  To impose a uniform shear, a slat (or boundary wall \X1  and \X2) at position $x$ moves with a velocity
\begin{equation}\label{eq:strain_rate}
    \mathbf{v}(x)=-\dot{\gamma} x\mathbf{u}_y
\end{equation}
where $\dot{\gamma}=2.1\times 10^{-3}s^{-1}$ is the shear strain rate. $\mathbf{u}_x$ and $\mathbf{u}_y$ are unit vectors in the horizontal and vertical directions on the figure.
The strain rate $\dot{\gamma}$ is held constant for all experiments. Walls \X3 and \X4 rotate in such a way that no slipping occurs at the junctions between them and the two other walls, consistent with the motions of the slats. The maximal static friction between a particle and the base slat is $0.0036\,$N, which is sufficient to entrain the rattler discs to the affine strain field but is negligible compared to the typical contact forces in jammed states. 
{\color{black} 
The base friction helps to form uniform shear-jammed states, which are presumably more stable than the states with shear bands that are formed using shear that is applied only from the boundary.}
More technical details on this device can be found in Ref.~\cite{ren2013_thesis,wang2018_phd}. In cyclic shearing, slight bending of the boundary walls 
and tiny slipping at the junctions when the direction of shear is changed lead to slightly asymmetric strain cycles, as discussed in detail in Appendix.~\ref{app_strain}.

The strain rate employed here is considered quasistatic for the following reasons. The two-dimensional inertial number $\mathcal{I}=\dot{\gamma}\sqrt{m_{\rm b}/p}$ is less than $10^{-4}$ for pressure $p$ larger than $1\,$N/m, which is the case for the states of interest, where $m_{\rm b}=1.47\times 10^{-3}\,$kg is the mass of a large disc. Also, for nearly stress-free states, a disc with a non-affine velocity $v_{na}=\dot{\gamma}d_{\rm s}$ becomes static within a characteristic time $t=v_{na}/\mu_{\rm base} g\approx10^{-5}$s that is much smaller than the macroscopic time scale $1/\dot{\gamma}\approx 0.5\times10^3$s. 
Consistent with these separations of time scales, we find that when the motor is stopped, we observe negligible relaxation of the particle positions. A detailed discussion on this type of relaxation is provided in Appendix~\ref{app_creep}.

A single run of the experiment begins with the preparation of a homogeneous, stress-free, random packing in a parallelogram frame chosen such that a forward shear strain $\gamma_{\rm I}$ will yield a rectangular configuration. We then perform three stages of quasistatic shearing, as depicted in Fig.~\ref{fig:procedure}(b): (1) the initial forward shear; (2) $N$ cycles of additional shear between $\pm \delta\gamma$; and (3) a large {\it reverse} shear of 
\textcolor{black}{$-\gamma_{\rm I}$} starting from the end of the last shear cycle. 
{\color{black} We note that initial shear with larger $\gamma_{\rm I}$ leads to original shear-jammed states with a more stable force network~\cite{Bi2011_nat,sarkar2013_prl,sarkar2015_pre,sarkar2016_pre,dong2018_prl,zhao2019_prl}. See supplementary video for an example evolution of the force network during the initial shear~\cite{Note1}.}
The number of cycles $N$ is typically 1500, but is larger for systems that take longer to relax, up to a maximum of 4800, and some data is collected for the $N=0$ case (i.e., no cyclic shear is applied between the initial forward shear and reverse shear). An typical stress-strain curve for $N=0$ is plotted 
{\color{black} using pink and blue triangles} in Fig.~\ref{fig:procedure}(c). The purple open circles, purple filled circles, and the light blue open squares in Fig.~\ref{fig:procedure}(c) are from a different experiment with $\gamma_{\rm I}=0.147$, $N = 1000$ and $\delta\gamma=0.95\%$, displaying the responses in the first shear cycle, the last shear cycle, and during the reverse shear process respectively. We study eleven values of $\gamma_{\rm I}$ ranging from 0\% to 21\% and eight values of $\delta\gamma$ ranging from 0\% to 1.54\% and indicated in Fig.~\ref{fig:procedure}(c).  These $\delta\gamma$ values are small compared to the strain interval needed to fully release the stress $\sigma_{xy}$ induced by the initial shear $\gamma_{\rm I}$. During the initial shear, both the pressure and shear stress of the system grow, showing no evidence of saturation even for the largest $\gamma_{\rm I}$ studied here. 

A high resolution camera (Canon 5D Mark II) accompanied by an automated imaging system with a polariscope is used to take images of the system. Details of the imaging system and post-processing procedures can be found in Ref.~\cite{wang2018_phd}. Capturing the images for one configuration takes about 30 seconds, during which the strain rate is set to zero. We measure particle positions and contact forces between particles. The particle centers are detected with an uncertainty around 0.01$d_{\rm s}$ using a Hough-transform technique~\cite{peng2007_JCISE,Ren2013_prl}. The vector contact forces between particles are measured using a well-known nonlinear fitting algorithm~\cite{majmudar2005_nature,Daniels2017_rsi,zadeh2019_gm}. Technical details of our implementation of the algorithm are provided in  Appendix.~\ref{sec_force_uncertainty}.  We also use an empirical intensity gradient method~\cite{Howell1999_prl,zadeh2019_gm,zhao2019_njp} to estimate the overall pressure {\it only} for data shown in Fig.~\ref{fig:condition_of_ultrastable}. The calibration of this method is described in Appendix~\ref{app_g2}. From the contact forces, we construct the stress tensor $\hat{\sigma}$ defined as in Refs.~\cite{christoffersen1981_jam,radjai1998_prl,Bi2011_nat} 
\begin{equation}\label{eq:stress_tensor}
    \hat{\sigma} = \frac{1}{S}\sum_{i=1}^{N_{\rm p}}\sum_{j=1}^{N_{\rm p}}\Theta_{ij}\mathbf{r}_{ij}\otimes \mathbf{f}_{ij},
\end{equation}
where $i$ and $j$ are indices of particles, $N_{\rm p}$ is the number of particles excluding the ones that belong to the boundary layer, $S$ is the sum of the Voronoi cell areas of these internal particles, $\mathbf{r}_{ij}$ is the displacement of the contact between particle $i$ and particle $j$ from the center of particle $i$, $\mathbf{f}_{ij}$ is the contact force exerted on particle $i$ by particle $j$, $\Theta_{ij}$ is a contact indicator with value 1 if particles $i$ and $j$ are in contact and 0 otherwise, and $\otimes$ denotes the vector outer product. We exclude contacts with fitted force magnitude less than 0.005N. From $\hat{\sigma}$ we calculate the pressure $p=-{\rm Tr}(\hat{\sigma})$ and the off-diagonal element $\sigma_{xy}$, which we term the shear stress in this work. We note that in most cases for our system the principal axes of $\hat{\sigma}$ lie in the $(1,1)$ (i.e., $x'$) and $(1,-1)$ (i.e., $y'$) directions, so that $|\sigma_{xy}|$ is equal to the second invariant.

\begin{figure*}[t]
    \centering
    \includegraphics[width = \textwidth]{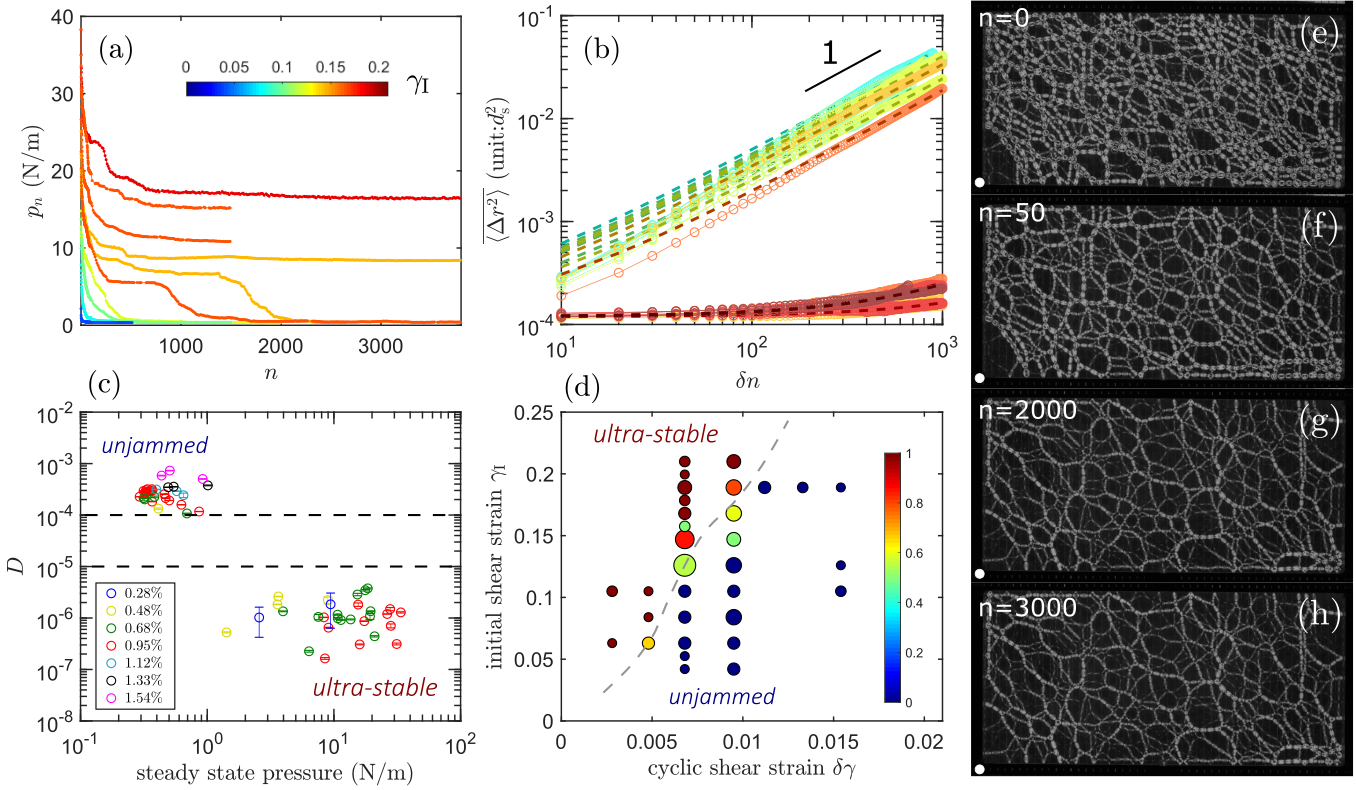}
    \caption{(a) The pressure measured after $n$ full shear cycles, $p_n$, as functions of shear cycle number $n$ for several example experiments with same cyclic shear strain amplitude $\delta\gamma=0.95\%$. The color of each curve in (a-b) labels the initial shear strain $\gamma_{\rm I}$ used to create the original shear jammed states, according to the colorbar in (a). (b) The mean square displacements of particle centers only in the steady-state regimes for several example experiments with same $\delta\gamma=0.95\%$. The unit is $d_{\rm s}^2$, where $d_{\rm s}$ is the diameter of our small disc. The dashed lines are linear fits on individual curves with form defined in Eq.~(\ref{eq:msd_fit}). (c) The diffusion coefficient $D$ obtained by the fits using Eq.~(\ref{eq:msd_fit}) for the steady states formed by different $\gamma_{\rm I}$ and $\delta\gamma$ plotted versus the pressure of these steady states. The color of the markers labels the $\delta\gamma$ value used to form these steady states. (d) The frequency of the observation of ultra-stable states under different control parameters $\gamma_{\rm I}$ and $\delta\gamma$ for about a hundred independent experiments each contains thousands of shear cycles. The size of each circle labels the number of experiments performed using same ($\delta\gamma,\gamma_{\rm I}$), with the smallest and the largest corresponds to 1 and 9 realizations. The color of each circle labels the number fraction of experiments that an ultra-stable state is observed, following the inserted colorbar. The dashed curve sketches the boundary that separates the ultra-stable states and the unjammed states as the outcome of cyclic shearing. (e-h) are snapshots of experiments showing the photoelastic patterns of an example experiment with $\gamma_{\rm I} = 0.147$ and $\delta\gamma = 0.95\%$ after 0, 50, 2000, and 3000 shear cycles respectively. Note that if the photoelastic patterns are the same for two packings then the positions of particles and the contact forces between particles are all the same. The white circles at the lower left corner of each panel have same diameter as our bigger disc (15.9~mm) and serve as scale bars for images in (e-h).}
    \label{fig:condition_of_ultrastable}
\end{figure*}

\section{Results}\label{sec_results}

\subsection{Formation of ultra-stable states}

We first study the parameter regime in which an ultra-stable state is formed.  The shear jammed states created by initial shear $\gamma_{\rm I}$ alone are unstable to cyclic shear for all the $\delta\gamma$ values that we studied. After a sufficiently large cycle numbers, however, the system evolves to one of two distinct types of steady state, depending on the values of $\gamma_{\rm I}$ and $\delta\gamma$. Generally speaking, ultra-stable states appear only for large $\gamma_{\rm I}$ and small $\delta\gamma$. If the system does not settle in an ultra-stable state, the accumulation of plastic deformations from cycle to cycle leads to a complete collapse of the packing. In this case, the material returns to a stress-free (unjammed) state. Figure~\ref{fig:condition_of_ultrastable}(a) shows the pressure measured after $n$ shear cycles for several example experiments with same $\delta\gamma=0.95\%$ but different $\gamma_{\rm I}$. It is clear that at long time limit the pressure can either reach a constant value, which indicates formation of an ultra-stable state, or drop to zero, which indicates a steady state consisting of a series of unjammed states. We note that the system may reach a metastable plateau before reaching the steady state, and the duration of a plateau could be very long near the transition. A detailed study of these plateaus is beyond the scope of the present paper.

Figure~\ref{fig:condition_of_ultrastable}(e-h) shows snapshots of the photoelastic patterns of the full system from an experiment with $\gamma_{\rm I}=0.147$ and $\delta\gamma = 0.95\%$ after 0, 50, 2000, and 3000 shear cycles. The photoelastic patterns are indistinguishable in (g) and (h).
{\color{black} 
The photoelastic patterns are indistinguishable in (g) and (h). We show only $n=2000$ and $n=3000$ states here due to limited space, but the strobed patterns are also indistinguishable for all cycles between $n=1000$ and $n = 3000$.}
Videos showing examples of the evolution of strobed states can be found in the Supplementary Material \footnote{See Supplementary Material for videos showing (1) the strobed states under cyclic shear in two cases where an ultra-stable state is formed, (2) the strobed states under cyclic shear in a case of relaxation to an unjammed state, (3) the stress relaxation over time for a shear-jammed state when the shear is suddenly stopped, 
{\color{black} and (4) the evolution of force network during an initial shear.}}.

Figure~\ref{fig:condition_of_ultrastable}(b) shows the mean-squared-displacement (MSD) of particle centers as functions of cycle number interval $\delta n$ in the steady-state regimes. We denote the MSD as $\overline{\langle \Delta r^2\rangle}$ where $\langle\cdot\rangle$ means averaging over particles and $\overline{\cdot}$ means averaging over starting points of the $\delta_n$ interval.  Figure~\ref{fig:condition_of_ultrastable}(b) shows experiments with same $\delta\gamma=0.95\%$ but different $\gamma_{\rm I}$, identified by the same colorbar used in Fig.~\ref{fig:condition_of_ultrastable}(a). The MSDs for ultra-stable states are quite small over a time scale of a thousand shear cycles, while the unjammed states show significant diffusive displacements. To quantify these observations, we fit each MSD curve to a linear form
\begin{equation}\label{eq:msd_fit}
    \overline{\langle\Delta r^2\rangle}(\delta n) = 4\times D\times4\delta\gamma\delta n + c_{\rm noise},
\end{equation}
where the diffusion coefficient $D$ is the fit parameter and $c_{\rm noise} = 1.2\times 10^{-4}\,d_s^2$ is the measured noise level of our particle center detection algorithm. Figure~\ref{fig:condition_of_ultrastable}(c) shows the fit results for $D$ as a function of the pressure of the corresponding steady state for all experiments, including different $\gamma_{\rm I}$ as well as  $\delta\gamma$. There are clearly two sets of steady states with $D$ separated by more than an order of magnitude. We note that all steady states with finite pressure have $D<10^{-5}\,d_{\rm s}^2$, corresponding to ultra-stable states. Moreover, the states with large $D$ all have nearly zero pressure, indicating unjammed states.

Figure~\ref{fig:condition_of_ultrastable}(d) shows the fraction of experiments performed with a given  ($\delta\gamma,\gamma_{\rm I}$) that produced an ultra-stable state, showing that such states occur only at small $\delta\gamma$ and high $\gamma_{\rm I}$. The dashed curve is a guide to the eye separating the unjammed and ultra-stable regimes. For $\delta\gamma$ and $\gamma_{\rm I}$ larger than the values shown in Fig.~\ref{fig:condition_of_ultrastable}(d), an out-of-plane instability prevents us from taking measurements on the quasi-2D granular material. The maximal $\delta\gamma$ for which we observe an ultra-stable state is 0.95\%. Figure~\ref{fig:condition_of_ultrastable}(c) suggests that there is a first-order dynamical transition when this boundary is crossed. {\color{black} We also observe that the number of cycles needed to reach a steady state peaks for parameter values near the phase boundary, reminiscent of a relevant dynamical transition observed in numerical glass~\cite{kawasaki2016_pre}.}
The location of the phase boundary in Fig.~\ref{fig:condition_of_ultrastable}(d) likely depends on several system parameters. Preliminary experiments suggest that for fixed $\delta\gamma$,  the initial strain required to produce ultra-stable states becomes smaller for higher volume fractions.  The effects of varying particle friction and bulk modulus have yet to be explored.

\subsection{Elasticity of the ultra-stable states}

\begin{figure}[th]
    \centering
    \includegraphics[width = \columnwidth]{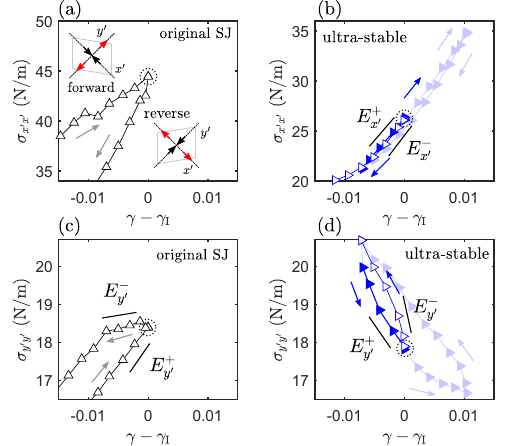}
    \caption{Measuring the slopes $E_{x'}^+$, $E_{x'}^-$, $E_{y'}^+$, and $E_{y'}^-$ near the states of interest from the stress-strain curves. The stresses $\sigma_{x'x'}$ and $\sigma_{y'y'}$ are stress tensor elements in the coordinate system $x'y'$ rotated $\pi/4$ from the original coordinate system $xy$ as shown in Fig.~\ref{fig:procedure}(a). 
    The $x'$ and $y'$ directions are the principal compression and dilation directions of the forward shear where $\gamma$ increases and they become the principal dilation and compression directions during reverse shearing where $\gamma$ decreases, as highlighted by the two sketches in (a). (a) and (c) plot the data measured during the initial shear and reverse shear near an example original shear-jammed (SJ) state formed by an initial shear with $\gamma_{\rm I} = 0.147$. The gray arrows mark the shearing directions. (b) and (d) plot an example ultra-stable state formed by applying cyclic shear with strain amplitude $\delta\gamma = 0.95\%$ on an original SJ state formed by an initial shear $\gamma_{\rm I} = 0.21$. In (b) and (d), filled triangles are data measured in the last shear cycle, and the open triangles are data measured in the reverse shearing process that follows (see Fig.~\ref{fig:procedure}(b)). The light blue data are those not used in calculating the slopes of interest. The arrows in (b) and (d) mark the shearing directions. In (a-d) the original SJ state and the ultra-stable state near which the slopes were measured are highlighted by the dashed black circles.}
    \label{fig:stress-strain-curves}
\end{figure}

The ultra-stable states behave much more like an ordinary elastic 
solid than the original 
shear-jammed states. We consider the mechanical response of the original states and the ultra-stable states to perturbations in the form of additional {\it forward} or {\it reverse} shear strain. We find that it is most useful to analyze the stress responses using a coordinate system $x'y'$ that is rotated by $\pi/4$ clockwise from the original coordinate system $xy$, as depicted in Fig.~\ref{fig:procedure}(a).The coordinates $x'$ and $y'$ then align with the principal compression and dilation directions of the initial simple shear deformation (see also the insert sketches in Fig.~\ref{fig:stress-strain-curves}(a)). We note that the $x'$ and $y'$ are also the principal directions of the stress tensors of the system in most cases. Thus, the stress tensor is diagonal in the rotated frame, its eigenvalues are $\sigma_{x'x'}$ and $\sigma_{y'y'}$, and we have $\sigma_{xy} = \frac{1}{2}(\sigma_{x'x'}-\sigma_{y'y'})$. Finally, the global shear stiffness can be decomposed as 
\begin{equation}\label{eq:slope_def}
    G = \frac{\partial  \sigma_{xy}}{\partial \gamma} = \frac{1}{2}\left(\frac{\partial \sigma_{x'x'}}{\partial \gamma} - \frac{\partial \sigma_{y'y'}}{\partial \gamma}\right) =\frac{E_{x'}-E_{y'}}{2},
\end{equation} 
where $E_{x'}= \partial \sigma_{x'x'}/\partial \gamma$ and $E_{y'}=\partial \sigma_{y'y'}/\partial \gamma$ are the contributions to the global shear stiffness from the responses along the two principal directions $x'$ and $y'$. We will denote the slopes $E_{x'}$ and $E_{y'}$ measured under forward or reverse shear using a superscript $^+$ or $^-$, respectively.

\begin{figure}[th]
    \centering
    \includegraphics[width = \columnwidth]{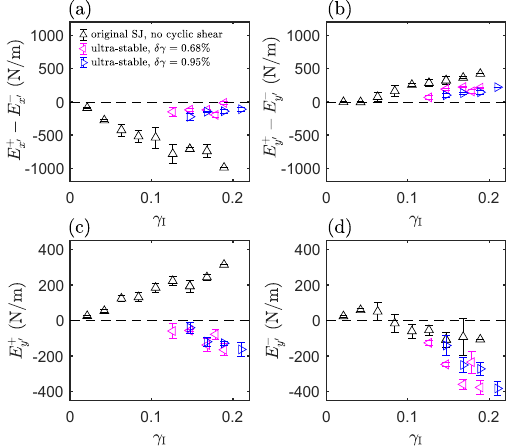}
    \caption{(a-b) The differences between the slopes $E_{x'}$ and $E_{y'}$ measured under the forward shear (denoted by the superscript $^+$) and under the reverse shear (denoted by the superscript $^-$) for the original shear-jammed (SJ) states and the ultra-stable states. (c-d) The slopes $E_{y'}^+$ and $E_{y'}^-$ for the states of interest.  The legend in (a) applies also for (b-d). In (a-d), the horizontal axis is the initial shear strain $\gamma_{\rm I}$ used to prepare the original SJ and the ultra-stable states. The cyclic strain amplitude used to prepare the ultra-stable states are given by the legend in (a).}
    \label{fig:moduli}
\end{figure}

Figure~\ref{fig:stress-strain-curves}(a) and (c) show the evolution of the stresses for an experiment with only initial shear and reverse shear, which is used to measure the responses of the original state created at $\gamma_{\rm I} = 0.147$. The slopes of the curves are measured by performing fits to 
the form
\begin{equation}\label{eq:fit_form}
\begin{split}
\sigma_{i'i'}(\gamma) & =E_{i'}(\gamma-\gamma_{\rm I})+E_{i',\rm nl}(\gamma-\gamma_{\rm I})^2+{\rm constant},
\end{split}
\end{equation}
where $E_{i'}$ and $E_{i',\rm nl}$ are fit parameters and $i$ can be either $x$ or $y$. The nonlinear terms are introduced for a better fit, but the values of $E_{i',\rm nl}$ are not of  interest for present purposes. We will denote the slopes $E_{x'}$ and $E_{y'}$ measured under forward or reverse shear using a superscript $^+$ or $^-$, respectively.

For the ultra-stable states, we measure
$E_{x'}$ and $E_{y'}$ by considering both the data in the last shear cycle and the reverse shear that follows (see Fig.~\ref{fig:procedure}(b) near $n=N$). In Fig.~\ref{fig:stress-strain-curves}(b) and (d), the last shear cycle data are plotted as filled triangles and the reverse shear data are plotted as the open triangles. The ultra-stable state being considered is highlighted by the black dashed circle. Again, 
data are fitted to Eq.~(\ref{eq:fit_form}) to obtain the slopes. The light blue data are not used in the fitting.

An ideal elastic medium 
would
display reversible stress-strain curves for which the slopes measured under forward or reverse shear are same. Figure~\ref{fig:moduli}(a) and (b) plot $E_{x'}^+-E_{x'}^-$ and $E_{y'}^+-E_{y'}^-$ for the original states and the ultra-stable states as functions of the initial shear strain $\gamma_{\rm I}$. Both differences are smaller for the ultra-stable states. Thus, we claim that the ultra-stable states behave more like an elastic material.

We note a nontrivial change in the material response along the $y'$ direction. Namely, $E_{y'}^+$ is positive and $E_{y'}^-$ is nearly zero for original states but both are negative for the ultra-stable states, as shown in Fig.~\ref{fig:moduli}(c) and (d).  Since $y'$ is the dilation principal direction of the forward shearing, negative $E_{y'}^+$ and $E_{y'}^-$ are expected for an ordinary elastic solid.


\begin{figure}[t]
    \centering
    \includegraphics[width = \columnwidth]{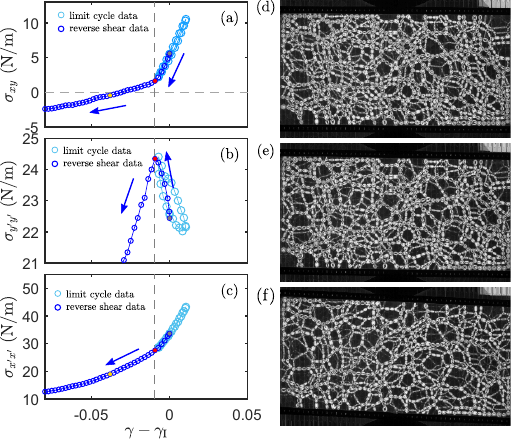}
    \caption{(a-c) The 
    blue circles are data under the reverse shear starting from an ultra-stable state formed by applying cyclic shear with strain amplitude 
    $\delta\gamma=0.95\%$
    on an original SJ state formed by 
    $\delta\gamma=0.189$. 
    The light blue circles show data in the 
    limit cycle. The arrows mark the direction of the reverse shearing. 
    Data in (a-c) are the shear stress $\sigma_{xy}$ and the two eigen values of the stress tensor $\sigma_{x'x'}$ and $\sigma_{y'y'}$. Note that there is $\sigma_{xy} = (\sigma_{x'x'}-\sigma_{y'y'}$)/2.
    The vertical dashed line marks the yielding transition.    (d-f) are snapshots of the system shown in (a-c) at $\gamma-\gamma_{\rm I}=0$ (the ultra-stable state),
    -0.0093 (at yielding point), and -0.0381
    respectively.
    The three states in (d-f) are marked by the circles filled by gray, red, and yellow color in (a-c).
    \label{fig:yield}}
\end{figure}

\subsection{Yielding of the ultra-stable states}

The ultra-stable states display a clear yielding transition  under reverse shearing. 
Unlike most cases of yielding in jammed systems, where plastic flow is induced by increasing shear stress, the yielding transition considered here is accompanied by a reduction of the shear stress. The yielding transition is evident in the evolution of the shear stress $\sigma_{xy}$
during reverse shearing. 
Figure~\ref{fig:yield}(a), (b), and (c) plot the
shear stress $\sigma_{xy}$ and  eigenvalues of the stress tensor $\sigma_{y'y'}$ and $\sigma_{x'x'}$
under reverse shear applied to an ultra-stable state prepared with $\gamma_{\rm I} = 
0.189$ and $\delta\gamma = 0.95\%$. The data from the limit cycle are also plotted. 
It is clear that the slope of $\sigma_{xy}$, $G$, changes sharply
when the strain reaches a threshold 
marked by the vertical dashed line, which is close to $\gamma_I - \delta\gamma$. At this point, $G$ drops to a much smaller value, indicating that the system suddenly becomes softer under reverse shear. We refer to this softening as a yielding transition. As $G=\frac{1}{2}(E_{x'}-E_{y'})$~(Eq.~\ref{eq:slope_def}), additional insight could be obtained by examining the slopes of the two eigenvalues. Remarkably, the slope of $\sigma_{y'y'}$ suddenly changes from negative to positive at the yielding point, while
the change in slope for 
$\sigma_{x'x'}$ is less dramatic. We also show some snapshots of the system across this yielding transition in Fig.~\ref{fig:yield}(d-f). 
{\color{black} Notably, both before and after yielding, there is always a strongly percolating force network. A quantitative characterization of the change in the force network during yielding is beyond the scope of the present paper.}

\section{Concluding remarks}\label{sec_discussion}

We have examined the stability of shear-jammed granular materials by applying small-amplitude shear cycles and monitoring all particle positions and contact forces. The observed emergent ultra-stable states exhibit qualitatively different responses to additional applied shear strain from those of the original shear jammed states. 

Our first major finding is the 
experimental observation of ultra-stable states for shear-jammed packings prepared by a large initial shear strain 
followed by small amplitude cyclic shear.
In an ultra-stable state all the particle positions, orientations, and contact forces become periodic, in strong contrast to the commonly encountered steady-state in which particles positions always rearrange~\cite{kou2017_nature,sun2020_prl} even though stress-strain curves may appear to be periodic~\cite{Ren2013_prl,leishangthem2017_natcom}. The existence of a limit cycle 
with periodic particle displacements and contact forces in frictional granular materials was first observed in numerical simulations reported by Royer et al.~\cite{Royer2015_pnas}. 
Interestingly, the limit cycle in our system can be induced by changing the initial shear strain $\gamma_{\rm I}$, a control parameter for the shear jamming process that is not considered in Ref.~\cite{Royer2015_pnas}.

Our second major finding is that cyclic shearing alters the mechanical properties of the shear jammed packing. In response to small perturbations, ultra-stable states
look more like ordinary elastic solids than do the original
shear jammed states.
{\color{black} Although elastic responses always dominate, there remains a measurable small hysteresis in the stress-strain curves that may come from reversible plastic events similar to those identified in Ref.~\cite{keim2014_prl}. }
This 
{\color{black} strongly}
elastic response extends to a strain near the cyclic shearing amplitude, where we have identified a yielding transition. The effect of friction on the mechanical properties of  limit cycles has been investigated in recent numerical simulations~\cite{Otsuki2021_epje}. However, these studies are focused on  packings above the isotropic jamming density, which are not as fragile as the shear-jammed states that we study.

\begin{figure}[t]
    \centering
    \includegraphics[width = \columnwidth]{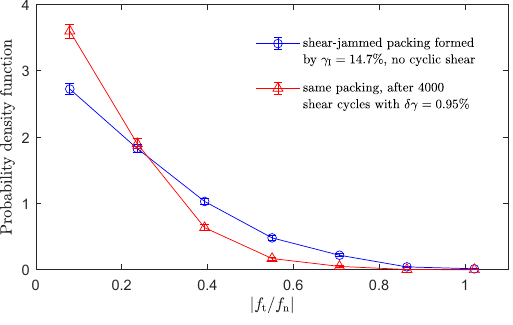}
    \caption{
    The distribution of the ratio between the tangential and normal components of the contact vector force in the original shear-jammed packing (blue circles) and the corresponding ultra-stable packing formed after cyclic shear (red triangles). 
    We exclude contacts with forces too weak to measure (around 0.05 N). {\color{black} Curves from experiments with different $\gamma_{\rm I}$ and $\delta\gamma$ are qualitatively similar.}
    }
    \label{fig:contact-network}
\end{figure}

In a preliminary attempt to discover the origin of ultra-stability in our system, 
we have measured the distribution over all contact forces of the ratio of tangential to normal force. We find that in the ultra-stable states the distribution has shifted away from the Coulomb limiting value, as shown in Fig.~\ref{fig:contact-network}. That is, the ultra-stable states have a smaller number of contacts with frictional components near the Coulomb limit.

The ultra-stable states observed in these frictional materials share some features with other jammed systems.  First, limit cycles with periodic particle positions have been found in experiments where particles are stabilized by electrostatic interaction and do not form contacts~\cite{keim2014_prl}, and also in experiments on foams~\cite{Lundberg2008_pre}. Our ultra-stable states may also share some features with the absorbing states in frictionless jammed solids~\cite{kawasaki2016_pre,ness2020_prl}.
In our system, however, contact friction is essential for mechanical stability as the measured mean contact number always remains below the frictionless isostatic number, whereas friction does not contribute to the stability of the states studied in Refs~\cite{keim2014_prl,Lundberg2008_pre}. Second, the yield strain for ultra-stable states is close to the cyclic strain amplitude used to prepare the state and thus constitutes a memory effect reminiscent of recent findings in other disordered systems~\cite{fiocco2013_pre,keim2019_rmp,arceri2021_pre2,Galloway2022_natphy}.  Notably, a recent experiment on a bubble raft suggested that the force network formed during cyclic shearing plays a decisive role in memory formation~\cite{Mukherji2019_prl}, though force chains could not be directly identified in that study Ref.~\cite{Mukherji2019_prl}. Our experimental system allows for a quantitative study of the force chains that stabilize the ultra-stable packings. Third, recent numerical and theoretical studies on model glasses show that cyclic shearing can mimic the role of annealing~\cite{das2018_arxiv,yeh2020_prl} and that different degrees of annealing may lead to distinct yielding behaviors~\cite{ozawa2018_pnas,yeh2020_prl}. 
Our observation that cyclic shearing changes the yielding properties of the original states thus provides an analogy to mechanical annealing in a frictional system.  Finally, 
while most previous experimental studies on limit cycles and memory effects examine 2D systems like ours, the phenomenon of shear jamming has been observed in 3D systems both in experiments~\cite{Peters2016_nat,han2019_prl} and simulations~\cite{luding2016_nature,Baity2017_jps,Jin2018_sa,Babu_2021Soft}. We hypothesize that ultra-stable states can also form in such 3D systems.

It is also worth emphasizing that not all shear-jammed states evolve to an ultra-stable state under cyclic shear. A large portion of seemingly strong shear-jammed packings completely collapse under quasistatic periodic shear with strain amplitude below 1\%. The weakening or breaking of a jammed, disordered structure under small mechanical perturbation is
reminiscent of recent findings such as the unjamming of frictionless discs/spheres~\cite{dagois2012_prl,das2020_pnas,Babu_2021Soft}, the liquefaction of soils~\cite{ming2021_pre,amini2021_GL}, the softening of colloidal gels~\cite{Gibaud2020_prx,Dages2021_jor}, and the reduction of viscosity of dense suspensions~\cite{lin2016_pnas,ness2018_sa,Sehgal2019_prl}.
We note that mechanical perturbations introduced by oscillatory shear~\cite{lin2016_pnas,ness2018_sa} or acoustic waves~\cite{Sehgal2019_prl} have been shown to enhance flow of a dense suspension by breaking the frictional contacts between particles. Our observation of  ultra-stable states suggests that applying perturbations to a strongly shear-jammed suspension may instead further stabilize the system, which suggests a new strategy for controlling the rheology of sheared granular suspensions. In our system, the change in nature of the steady states induced by cyclic shear from unjammed (for large amplitude cycles or small initial shear strain) to ultra-stable (for smaller amplitudes or larger initial shear strain) shows features of a first-order dynamical phase transition, including a sudden jump in the diffusion coefficient, reminiscent of a similar transition in model glasses~\cite{kawasaki2016_pre}. The appearance of the long-lived meta-stable plateaus near the transition is an intriguing phenomenon calling for further study.

Our results have broad implications for the handling of granular materials. For example, understanding the stability of shear jammed states may help design protocols and devices to enhance flow or avoid blockages in sheared dense suspensions. In addition, our experimental observations provide significant clues for the development of theories of  protocol-dependent mechanical properties of granular systems.

\end{CJK*}
\begin{acknowledgments}
We thank Shuai Zhang, Yinqiao Wang, Ryan Kozlowski, Jonathan Bar\'{e}s, Francesco Arceri, Hanqing Liu, Karen E. Daniels, Raphael Blumenfeld, and Patrick Charbonneau for helpful discussions. This work was primarily supported by NSF grant DMR-1809762, BC was supported by NSF grants CBET-1916877, and CMMT-2026834, and BSF-
2016188. HZ thanks the support from the Fundamental Research Funds for the Central Universities No. 22120210143.
\end{acknowledgments}

\appendix

\section{Contact force law on a single disc}\label{app_deflection}

\begin{figure}[t]
    \centering
    \includegraphics[width = \columnwidth]{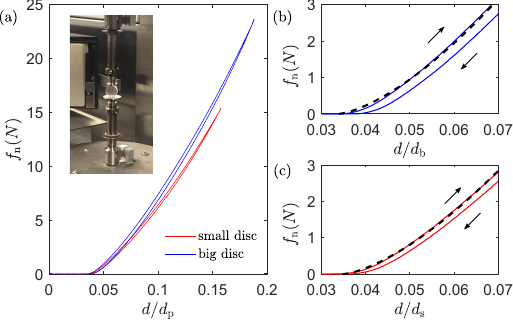}
    \caption{(a) Main panel: the normal contact force $f_{\rm n}$ experienced by a small disc (red) and a big disc (blue) measured by the Micro-Strain Analyzer as functions of distance $d$ moved by its upper arm scaled by the diameter of the disc. Note that $d_{\rm p}=d_{\rm s}$ for the small disc and $d_{\rm p}=d_{\rm b}$ for the big disc. 
    The discs experience a compression-decompression load cycle, which leads to weak but noticeable hysteresis. The insert panel shows a picture of the loading part of the TA Instruments RSA III Micro-Strain Analyzer, taken from Ref.~\cite{zhao2020_phd} with permission.
 (b) and (c) show zoom-in versions of same data as in (a), where the evolution direction of the system is indicated by the arrows. The black dashed curves in (b) and (c) are fit results using Eq.~(\ref{eq:sm:forcelaw}) for the big and small disc respectively.}
    \label{fig:sm:forcelaw}
\end{figure}

We measure the relationship between the contact force and the deformation of a single disc through diametric compression and decompression tests using a TA Instruments RSA III Micro-Strain Analyzer. The instrument measures the distance travelled by the upper arm, $d$, and the normal contact force magnitude, $f_{\rm n}$. A picture of the loading part of this instrument is shown in Fig.~\ref{fig:sm:forcelaw}(a), taken from Ref.~\cite{zhao2020_phd} with permission. The relation between $d$, rescaled by the diameter of the corresponding disc, and $f_{\rm n}$ are plotted in Fig.~\ref{fig:sm:forcelaw}(a) for a small disc and a big disc. Weak hysteresis can be observed, reflecting the viscoelastic nature of the polyurethane discs. Figure~\ref{fig:sm:forcelaw} plots the compression and decompression curves for a big and a small disc. The probe moves at a constant speed of 0.03 mm/s, results in a strain rate about 0.25\% per second for a single disc, similar to the shear rate used in the experiment.

Most contact forces in this work are below 3~N. We show that the Hertzian contact force law is a reasonable approximation in this regime. Figure~\ref{fig:sm:forcelaw}(b) and (c) plot same data in (a) but only near the touching point between the probe arm and the corresponding disc. For both discs, we fit data between 0.01~N and 3~N obtrained during the compression process to the following form 
\begin{equation}\label{eq:sm:forcelaw}
        f_{\rm n}=\frac{\epsilon}{d_{\rm p}}(\frac{d-d_{\rm c}}{d_{\rm p}})^{\frac{3}{2}},
\end{equation}
and get $\epsilon=5.46\pm 0.06$ N$\cdot$m and $d_{\rm c}/d_{\rm s}=0.0347\pm 0.0002$ for the small disc and $\epsilon = 7.25\pm 0.08$ N$\cdot$m and $d_{\rm c}/d_{\rm b}=0.0337\pm 0.0002$ for the big disc. Note that $d_{\rm p}$ is the diameter of the disc being considered. $d_{\rm c}$ is the point when the upper arm of the Micro-Strain Analyzer just touches the particle. The fit results are plotted as the black dashed curves in Fig.~\ref{fig:sm:forcelaw}(b) and (c). We note that Eq.~(\ref{eq:sm:forcelaw}) slightly overestimates small forces, for which a quadratic form appears to fit better. The closed-form solution for this problem is given in Ref.~\cite{norden1973_report}.   

\section{Measurement of the global shear strain using particle displacements and minor deviations from uniform shear near the onset of shear reversal}\label{app_strain}

\begin{figure}[!ht]
    \centering
    \includegraphics[width = \columnwidth]{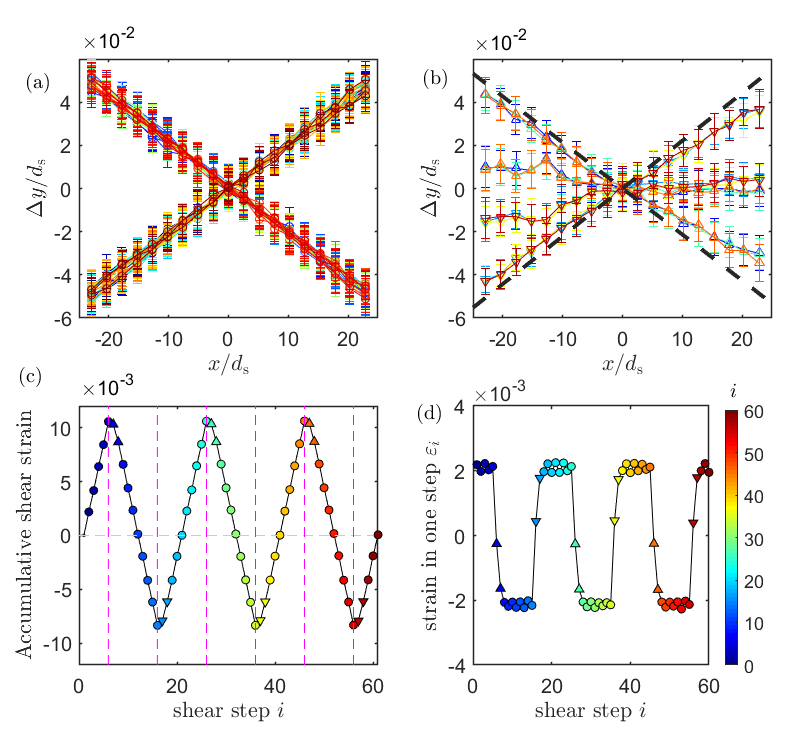}
    \caption{(a) Particle center displacements along the y direction as functions of their x positions between consecutive quasi-static shear steps that show nice linear relation as expected for a uniform shear. The coordinate system is plotted in Fig.~\ref{fig:procedure}(a). Each curve corresponds to averaged data for one shear step (from $i$th to $i+1$th step, with $i$ labeled based on colorbar to the right of subfigure (d)) with a bin width 2.5 $d_{\rm s}$, the diameter of a small disc. (b) Particle displacements from the abnormal steps following the change of driving directions. The curves are measured from shear steps with same type of marker shape and color shown in (c) and (d). In (a) and (b), error bars show standard deviation obtained from the averaging process, which is about $0.01d_{\rm s}$, near the limit of our center detection precision. In (b) the two dashed black lines show linear fit results from data in (a) that are under forward and reverse shear. (c) The accumulative shear strain starting from a state right before one complete shear cycle. In both (c) and (d) the triangles show abnormal steps and the circles show normal steps. (d) The strain caused by one shear step as defined in Eq.~(\ref{eq:def_strain_correction}).}
    \label{fig:sm:strain_correction}
\end{figure}

We find that the boundary walls and the base slats impose a uniform simple shear strain field very well in most of the cases both during both forward shear and reverse shear processes, as plotted in Fig.~\ref{fig:sm:strain_correction}(a) for an example cyclic shear experiment. The shear cell is driven by a stepper motor, which precisely controls the position and motion of the left end of wall \X4~\cite{ren2013_thesis,wang2018_phd}. A shear step here means that the left end of wall \X4 moves a fixed distance 1~mm. When the direction of driving is switched, the release and rebuild of a small elastic deformation of the two long and thin confining aluminum walls \X3 and \X4, and small relative motion at the joints between different walls and slats  may both cause minor deviations from the expected uniform shear strain field. Such a deviation is evidenced from the particle displacements measured at the two shear steps right after the change of shear direction as plotted in Fig.~\ref{fig:sm:strain_correction}(b). The actual shear strain experienced by the system in these two steps is smaller than what is expected from assuming a uniform strain field given the well-controlled motion of the left end of wall \X4.

In this work we measure the actual global shear strain of the material using the particle displacements. Given the bin-averaged displacement field during $i$~th shear step such as the curves in Fig.~\ref{fig:sm:strain_correction}(a) and (b), the global shear strain caused by this shear step, $\varepsilon_i$, is defined as 
\begin{equation}\label{eq:def_strain_correction}
    \varepsilon_i = \frac{1}{N_{\rm bin}-2}\sum_{k=2}^{N_{\rm bin}-1}\frac{\Delta y_{k+1}-\Delta y _{k-1}}{5d_{\rm s}},
\end{equation}
where $5d_{\rm s}$ is the twice of the bin size, $N_{\rm bin} = 19$ is the number of bins used to calculate the averaged data. Then the accumulative shear strain $\gamma$ is calculated by doing a summation of $\varepsilon_i$ over consecutive shear steps. Figure~\ref{fig:sm:strain_correction}(c) and (d) plot the accumulative shear strain and shear strain caused by each shear step for an example system experiencing three shear cycles. In Fig.~\ref{fig:sm:strain_correction}(c) purple dashed lines show times at which the direction of shear is changed. We see that the shear cycle is slightly non-symmetric: the minimal accumulative strain is -0.0085, yet the maximal one is 0.0105. We take the average between the absolute value of the two, $0.0095$, as the strain amplitude for such a driving. The uncertainty of this strain amplitude measurement is around 0.001.

We note that this small deviation near strain reversal does not affect any of our conclusions: the fits used to measure elastic responses do not use any data points right after shear reversal. Therefore these measurements indeed reflect properties of a uniformly deformed material. The exponential fits used to measure the plastic responses during shear reversal include shear steps with small deviations from the uniform strain field. However, the deviation effect is the same for both the ultra-stable states and the original shear jammed states. Thus, such an effect would not affect any claim based on the comparison between them.

\section{Contact force measurements using photoelasticity} \label{sec_force_uncertainty}

We introduce the implementation of the non-linear fitting algorithm we used to measure the contact forces and then report our estimation of the uncertainties of the measurement. We use a Matlab implementation adapted from PEGS~\cite{Daniels2017_rsi} with several modifications that improve the quality of the solution for the large forces, including (1) a neural network trained to give initial guesses, (2) the use of reaction forces and/or forces at an earlier strain step to refit particles with a large error at current step, and (3) manually supplied initial guesses determined using an interactive graphic interface for rare cases.  We enforce force and torque balance constraints on individual particles except for rare cases, and use the deviation between action and reaction forces at the contacts to estimate uncertainties of the measurements. In rare cases when some particles are bearing extremely large forces, we do not enforce force and torque balance. Instead we let the algorithm minimize the intensity differences and the residual net force and torque together. Such a method typically leads to smaller intensity differences between the reconstructed photoelastic images and the raw experimental images, and the residual net force and torque are usually negligible. The stress-optic coefficient of the particles is $F_{\rm \sigma}=157$, defined as in Ref.~\cite{Daniels2017_rsi} and measured using a technique detailed in Chapter~3.3.2 of Ref.~\cite{zhao2020_phd}. The feasible contact positions are detected for each disc before fitting. If the distance between the centers of two particles is less than 1.03 times the sum of their radii, we register a feasible contact and find the fitted contact force carried by it for both of the two contacting discs. If for both discs the magnitude of the contact force is less than $0.005$~N, we drop this contact and fit the discs again using only the remaining contacts. Such a process is repeated until all the remaining contacts are bearing forces whose magnitudes are larger than $0.005$~N.

\begin{figure}[!t]
    \centering
    \includegraphics[width =\columnwidth]{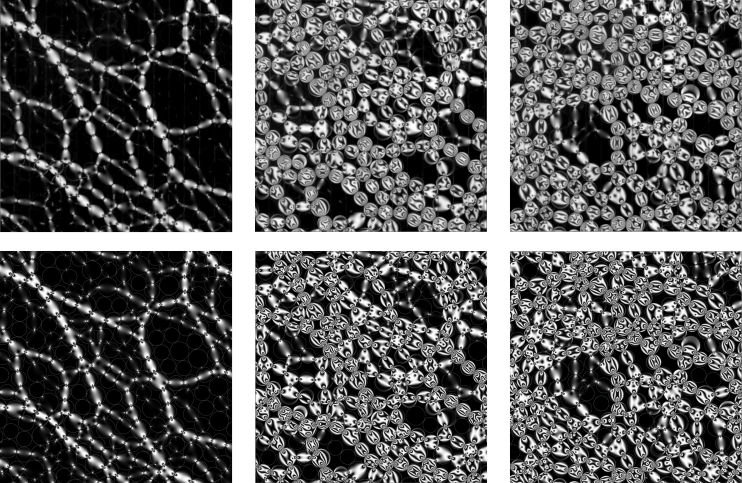}
    \caption{Comparison between the the experimental images taken through a polariscope (top row) and the reconstructed images based on contact force solutions (bottom row) for three example packings. The visual match between the phtoelastic patterns is an evidence of finding the global minimum of the error function~\cite{zadeh2019_gm,Daniels2017_rsi,zhao2020_phd}. The pressure of the packing from left to right are 6.8 N/m, 42.4 N/m, and 53.8 N/m.}
    \label{fig:sm:inverse:compare_patterns}
\end{figure}

\begin{figure}[!t]
    \centering
    \includegraphics[width = \columnwidth]{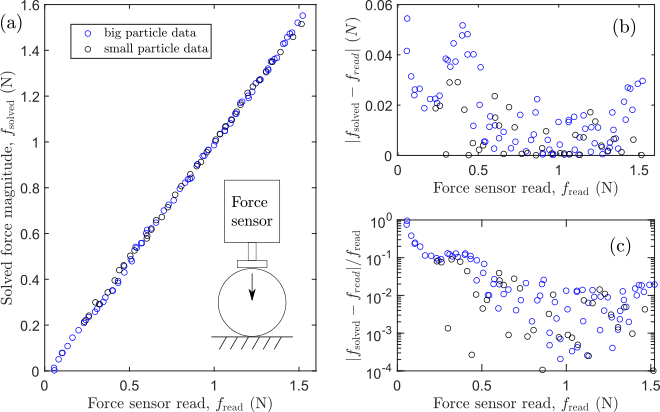}
    \caption{(a) Solved contact force magnitude from the fitting algorithm compared to the read from a commercial force sensor in a diametric loading test on a single disc. (b) and (c) plots the absolute error and the relative error of the solved forces.}
    \label{fig:sm:inverse:calibration}
\end{figure}

We show that our fitting algorithm finds the global minimum of the error function~\cite{majmudar2005_nature,Daniels2017_rsi,zadeh2019_gm} by showing examples of experimental images and reconstructed images in Fig.~\ref{fig:sm:inverse:compare_patterns}. We show the number of the photoelastic fringes and their overall shapes are very well reproduced even for packings with large pressure. We then compare quantitatively the forces solved by the fitting algorithm and the force measured from a commercial force sensor for a particle under diametric loading (Fig.~\ref{fig:sm:inverse:calibration}). In such a test, the absolute error is below 0.05 N, and the relative error is less than 10\% for forces larger than $\sim 0.1$ N.

\section{Intensity gradient method to estimate pressure for states formed during relaxation}\label{app_g2}

The pressure data shown in Fig.~\ref{fig:condition_of_ultrastable} are not calculated from contact forces but estimated using an empirical method called intensity gradient method. Such a method is first introduced in Ref.~\cite{Howell1999_prl} and more information can be found in recent reviews such as Ref.~\cite{Daniels2017_rsi} and Ref.~\cite{zadeh2019_gm}. It is a good particle-scale indicator for pressure only when the tangential forces are small compared to normal forces~\cite{zhao2019_njp}. In our experiments we find that contacts bearing tangential forces comparable to normal forces are rare, suggesting the applicability of this method for the system-averaged pressure. 

The mean intensity gradient of the packing, denoted as $g_2$ here, is defined as
\begin{equation}
\begin{split}
 g_2 =& 10^4 \frac{1}{N_{\rm p}}\sum_{i=1}^{N_{\rm p}}\frac{1}{N_{{\rm pixel},i}}\sum_{\substack{{\rm pixel} (i,j)\\{\rm in~disc}~i}}\frac{1}{4}\Big( (\frac{I_{i+1,j}-I_{i-1,j}}{2})^2\\
 &+(\frac{I_{i,j+1}-I_{i,j-1}}{2})^2+(\frac{I_{i+1,j+1}-I_{i-1,j-1}}{2\sqrt{2}})^2\\
 &+(\frac{I_{i+1,j-1}-I_{i-1,j+1}}{2\sqrt{2}})^2 \Big),
 \end{split}
\end{equation}
where $I_{i,j}$ is the rescaled intensity in the ($i,j$) pixel of the polarized image, which ranges from 0 to 1. $N_{{\rm pixel},i}$ is the number of pixels in $i$th disc. The prefactor $10^4$ is introduced in order to adjust $g_2$ value to be at a similar order of $p$.

\begin{figure}[t]
    \centering
    \includegraphics[width = \columnwidth]{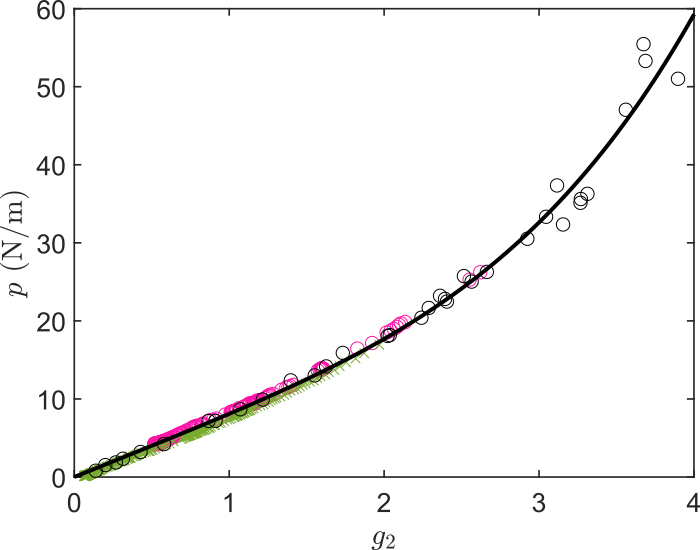}
    \caption{The pressure $p$ calculated from contact forces solved using the non-linear fitting algorithm plotted as function of $g_2$. The black circles are original states and the colored data points are states formed during cyclic shearing for two different experiments. The black curve plots Eq.~(\ref{eq:sm:g2}).}
    \label{fig:sm:g2}
\end{figure}

We calibrate the relation between the system-averaged pressure $p$ and $g_2$ using a set of original states where the values of $p$ are calculated using the contact forces solved by the nonlinear fitting algorithm. The relation between $p$ and $g_2$ for these states are plotted using black circles in Fig.~\ref{fig:sm:g2}. We fit these data using 
\begin{equation}\label{eq:sm:g2}
    p=ag_2+bg_2^4,
\end{equation}
which gives $a=8.0\pm 0.6$ N/m and $b=0.11\pm 0.02$ N/m. We have solved the contact forces for all states formed in two example cyclic shearing experiments and plot their $p$ and $g_2$ in Fig.~\ref{fig:sm:g2} as well, showing that these states follow same $p(g_2)$ relation as the original states. Thus, the pressure data shown in Fig.~\ref{fig:condition_of_ultrastable} are calculated using Eq.~(\ref{eq:sm:g2}).

\section{Stress relaxation of a shear-jammed state without oscillatory shear}\label{app_creep}

We argued in Sec.~\ref{sec_methods} that the shear rate used in this work is in the quasi-static regime. Here we clarify the precise meaning of this claim, which rests on making a distinction between time scales associated with granular dynamics and with material or contact aging.  When we measure stress as function of time right after stopping the initial shear that generates an initial shear-jammed state, we find that the rate of stress relaxation is very small and that there is no change in the contact network structure, indicating that the small stress change is principally due to the relaxation of the polymer material of the discs or the aging of frictional contacts.  On the time scale relevant for the rearrangement of the contact network, the relaxation is negligible; the contact network is already in force and torque balance when the shear is stopped. 

\begin{figure}[!t]
    \centering
    \includegraphics[width = \columnwidth]{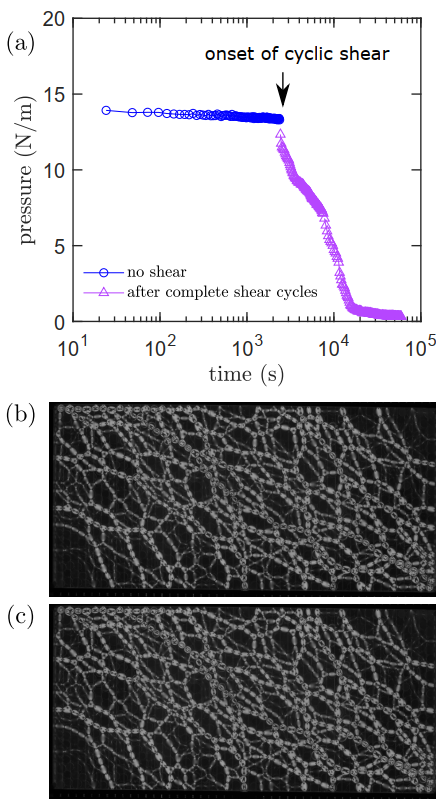}
    \caption{(a) The pressure evolution of an original shear-jammed state formed at time 0s by an initial shear strain $\gamma_{\rm I}=0.126$. From 0s to 2376s there is no deformation applied to the system, and the pressure is measured per 24s (blue circles). Starting from 2376s, we start cycle shear and record pressure after complete shear cycles (purple triangles). (b) and (c) are snapshots of the system at time 0s and time 2376s. See supplementary video 3~\cite{Note1} for a video of this relaxation process with a zoom-in window.}
    \label{fig:app_creep}
\end{figure}

In the example experiment shown in Fig.~\ref{fig:app_creep}, a shear-jammed state is formed by applying $\gamma_{\rm I} = 0.126$ initial shear to an unjammed state. Right after stopping the initial shear, we measure the pressure every 24 seconds, indicated by the blue circles in Fig.~\ref{fig:app_creep}(a). No additional shear is applied to the system until time 2376$\,$s. 
The pressure decays about 1.3 N/m during this time.
The snapshots of the force networks at times 0s and 2376s are shown in Fig.~\ref{fig:app_creep}(b) and (c). The two images look same, suggesting the force network is already in a force and torque balance at time 0s, which  indicates that the shear rate used to from the jammed state at time 0s is in the quasi-static regime. The slight difference between the two images can be better seen in supplementary video 3~\cite{Note1}. The ultra-stable states relax at a similarly slow rate when the boundary frame is held fixed.

Figure~\ref{fig:app_creep}(a) shows the pressure after complete shear cycles when we turn on the oscillatory shear with strain amplitude $\delta\gamma=0.95\%$ after 2376s.  The relaxation caused by oscillatory shear (applied at the same rate as the initial quasi-static shear) is much more significant than the relaxation of stress without any applied deformation. In addition, this quasi-static oscillatory shear changes the contact network structure, indicating that there is a strong separation between the material aging and granular dynamics time scales. Such a qualitative difference is easily seen when comparing supplementary video 3 to supplementary video 1 or 2~\cite{Note1}.  Thus it is reasonable to classify the applied shear as quasi-static for the purposes of analyzing the granular dynamics.





\bibliography{main}

\begin{thebibliography}{75}%
\makeatletter
\providecommand \@ifxundefined [1]{%
 \@ifx{#1\undefined}
}%
\providecommand \@ifnum [1]{%
 \ifnum #1\expandafter \@firstoftwo
 \else \expandafter \@secondoftwo
 \fi
}%
\providecommand \@ifx [1]{%
 \ifx #1\expandafter \@firstoftwo
 \else \expandafter \@secondoftwo
 \fi
}%
\providecommand \natexlab [1]{#1}%
\providecommand \enquote  [1]{``#1''}%
\providecommand \bibnamefont  [1]{#1}%
\providecommand \bibfnamefont [1]{#1}%
\providecommand \citenamefont [1]{#1}%
\providecommand \href@noop [0]{\@secondoftwo}%
\providecommand \href [0]{\begingroup \@sanitize@url \@href}%
\providecommand \@href[1]{\@@startlink{#1}\@@href}%
\providecommand \@@href[1]{\endgroup#1\@@endlink}%
\providecommand \@sanitize@url [0]{\catcode `\\12\catcode `\$12\catcode
  `\&12\catcode `\#12\catcode `\^12\catcode `\_12\catcode `\%12\relax}%
\providecommand \@@startlink[1]{}%
\providecommand \@@endlink[0]{}%
\providecommand \url  [0]{\begingroup\@sanitize@url \@url }%
\providecommand \@url [1]{\endgroup\@href {#1}{\urlprefix }}%
\providecommand \urlprefix  [0]{URL }%
\providecommand \Eprint [0]{\href }%
\providecommand \doibase [0]{https://doi.org/}%
\providecommand \selectlanguage [0]{\@gobble}%
\providecommand \bibinfo  [0]{\@secondoftwo}%
\providecommand \bibfield  [0]{\@secondoftwo}%
\providecommand \translation [1]{[#1]}%
\providecommand \BibitemOpen [0]{}%
\providecommand \bibitemStop [0]{}%
\providecommand \bibitemNoStop [0]{.\EOS\space}%
\providecommand \EOS [0]{\spacefactor3000\relax}%
\providecommand \BibitemShut  [1]{\csname bibitem#1\endcsname}%
\let\auto@bib@innerbib\@empty
\bibitem [{\citenamefont {Jaeger}\ \emph {et~al.}(1996)\citenamefont {Jaeger},
  \citenamefont {Nagel},\ and\ \citenamefont {Behringer}}]{Jeager1996_rmp}%
  \BibitemOpen
  \bibfield  {author} {\bibinfo {author} {\bibfnamefont {H.~M.}\ \bibnamefont
  {Jaeger}}, \bibinfo {author} {\bibfnamefont {S.~R.}\ \bibnamefont {Nagel}},\
  and\ \bibinfo {author} {\bibfnamefont {R.~P.}\ \bibnamefont {Behringer}},\
  }\bibfield  {title} {\emph {\bibinfo {title} {Granular solids, liquids, and
  gases}},\ }\href {https://doi.org/10.1103/RevModPhys.68.1259} {\bibfield
  {journal} {\bibinfo  {journal} {Rev. Mod. Phys.}\ }\textbf {\bibinfo {volume}
  {68}},\ \bibinfo {pages} {1259} (\bibinfo {year} {1996})}\BibitemShut
  {NoStop}%
\bibitem [{\citenamefont {de~Gennes}(1999)}]{deGennes1999_rmp}%
  \BibitemOpen
  \bibfield  {author} {\bibinfo {author} {\bibfnamefont {P.~G.}\ \bibnamefont
  {de~Gennes}},\ }\bibfield  {title} {\emph {\bibinfo {title} {Granular matter:
  a tentative view}},\ }\href {https://doi.org/10.1103/RevModPhys.71.S374}
  {\bibfield  {journal} {\bibinfo  {journal} {Rev. Mod. Phys.}\ }\textbf
  {\bibinfo {volume} {71}},\ \bibinfo {pages} {S374} (\bibinfo {year}
  {1999})}\BibitemShut {NoStop}%
\bibitem [{\citenamefont {Liu}\ and\ \citenamefont
  {Nagel}(1998)}]{Liu1998_nature}%
  \BibitemOpen
  \bibfield  {author} {\bibinfo {author} {\bibfnamefont {A.~J.}\ \bibnamefont
  {Liu}}\ and\ \bibinfo {author} {\bibfnamefont {S.~R.}\ \bibnamefont
  {Nagel}},\ }\bibfield  {title} {\emph {\bibinfo {title} {Jamming is not just
  cool any more}},\ }\href {https://doi.org/10.1038/23819} {\bibfield
  {journal} {\bibinfo  {journal} {Nature}\ }\textbf {\bibinfo {volume} {396}},\
  \bibinfo {pages} {21} (\bibinfo {year} {1998})}\BibitemShut {NoStop}%
\bibitem [{\citenamefont {O'Hern}\ \emph {et~al.}(2003)\citenamefont {O'Hern},
  \citenamefont {Silbert}, \citenamefont {Liu},\ and\ \citenamefont
  {Nagel}}]{OHern2003_pre}%
  \BibitemOpen
  \bibfield  {author} {\bibinfo {author} {\bibfnamefont {C.~S.}\ \bibnamefont
  {O'Hern}}, \bibinfo {author} {\bibfnamefont {L.~E.}\ \bibnamefont {Silbert}},
  \bibinfo {author} {\bibfnamefont {A.~J.}\ \bibnamefont {Liu}},\ and\ \bibinfo
  {author} {\bibfnamefont {S.~R.}\ \bibnamefont {Nagel}},\ }\bibfield  {title}
  {\emph {\bibinfo {title} {Jamming at zero temperature and zero applied
  stress: The epitome of disorder}},\ }\href
  {https://doi.org/10.1103/PhysRevE.68.011306} {\bibfield  {journal} {\bibinfo
  {journal} {Physical Review E}\ }\textbf {\bibinfo {volume} {68}},\ \bibinfo
  {pages} {011306} (\bibinfo {year} {2003})}\BibitemShut {NoStop}%
\bibitem [{\citenamefont {Majmudar}\ \emph {et~al.}(2007)\citenamefont
  {Majmudar}, \citenamefont {Sperl}, \citenamefont {Luding},\ and\
  \citenamefont {Behringer}}]{Majmudar2007_prl}%
  \BibitemOpen
  \bibfield  {author} {\bibinfo {author} {\bibfnamefont {T.~S.}\ \bibnamefont
  {Majmudar}}, \bibinfo {author} {\bibfnamefont {M.}~\bibnamefont {Sperl}},
  \bibinfo {author} {\bibfnamefont {S.}~\bibnamefont {Luding}},\ and\ \bibinfo
  {author} {\bibfnamefont {R.~P.}\ \bibnamefont {Behringer}},\ }\bibfield
  {title} {\emph {\bibinfo {title} {Jamming transition in granular systems}},\
  }\href {https://doi.org/10.1103/PhysRevLett.98.058001} {\bibfield  {journal}
  {\bibinfo  {journal} {Phys. Rev. Lett.}\ }\textbf {\bibinfo {volume} {98}},\
  \bibinfo {pages} {058001} (\bibinfo {year} {2007})}\BibitemShut {NoStop}%
\bibitem [{\citenamefont {Liu}\ and\ \citenamefont
  {Nagel}(2010)}]{Liu2010_arcmp}%
  \BibitemOpen
  \bibfield  {author} {\bibinfo {author} {\bibfnamefont {A.~J.}\ \bibnamefont
  {Liu}}\ and\ \bibinfo {author} {\bibfnamefont {S.~R.}\ \bibnamefont
  {Nagel}},\ }\bibfield  {title} {\emph {\bibinfo {title} {The jamming
  transition and the marginally jammed solid}},\ }\href
  {https://doi.org/10.1146/annurev-conmatphys-070909-104045} {\bibfield
  {journal} {\bibinfo  {journal} {Annual Review of Condensed Matter Physics}\
  }\textbf {\bibinfo {volume} {1}},\ \bibinfo {pages} {347} (\bibinfo {year}
  {2010})}\BibitemShut {NoStop}%
\bibitem [{\citenamefont {Bi}\ \emph {et~al.}(2011)\citenamefont {Bi},
  \citenamefont {Zhang}, \citenamefont {Chakraborty},\ and\ \citenamefont
  {Behringer}}]{Bi2011_nat}%
  \BibitemOpen
  \bibfield  {author} {\bibinfo {author} {\bibfnamefont {D.}~\bibnamefont
  {Bi}}, \bibinfo {author} {\bibfnamefont {J.}~\bibnamefont {Zhang}}, \bibinfo
  {author} {\bibfnamefont {B.}~\bibnamefont {Chakraborty}},\ and\ \bibinfo
  {author} {\bibfnamefont {R.~P.}\ \bibnamefont {Behringer}},\ }\bibfield
  {title} {\emph {\bibinfo {title} {Jamming by shear}},\ }\href
  {https://doi.org/10.1038/nature10667} {\bibfield  {journal} {\bibinfo
  {journal} {Nature}\ }\textbf {\bibinfo {volume} {480}},\ \bibinfo {pages}
  {355} (\bibinfo {year} {2011})}\BibitemShut {NoStop}%
\bibitem [{\citenamefont {Behringer}\ and\ \citenamefont
  {Chakraborty}(2018)}]{behringer2018_rpp}%
  \BibitemOpen
  \bibfield  {author} {\bibinfo {author} {\bibfnamefont {R.~P.}\ \bibnamefont
  {Behringer}}\ and\ \bibinfo {author} {\bibfnamefont {B.}~\bibnamefont
  {Chakraborty}},\ }\bibfield  {title} {\emph {\bibinfo {title} {The physics of
  jamming for granular materials: a review}},\ }\href
  {https://doi.org/10.1088/1361-6633/aadc3c} {\bibfield  {journal} {\bibinfo
  {journal} {Reports on Progress in Physics}\ }\textbf {\bibinfo {volume}
  {82}},\ \bibinfo {pages} {012601} (\bibinfo {year} {2018})}\BibitemShut
  {NoStop}%
\bibitem [{\citenamefont {Dagois-Bohy}\ \emph {et~al.}(2012)\citenamefont
  {Dagois-Bohy}, \citenamefont {Tighe}, \citenamefont {Simon}, \citenamefont
  {Henkes},\ and\ \citenamefont {van Hecke}}]{dagois2012_prl}%
  \BibitemOpen
  \bibfield  {author} {\bibinfo {author} {\bibfnamefont {S.}~\bibnamefont
  {Dagois-Bohy}}, \bibinfo {author} {\bibfnamefont {B.~P.}\ \bibnamefont
  {Tighe}}, \bibinfo {author} {\bibfnamefont {J.}~\bibnamefont {Simon}},
  \bibinfo {author} {\bibfnamefont {S.}~\bibnamefont {Henkes}},\ and\ \bibinfo
  {author} {\bibfnamefont {M.}~\bibnamefont {van Hecke}},\ }\bibfield  {title}
  {\emph {\bibinfo {title} {Soft-sphere packings at finite pressure but
  unstable to shear}},\ }\href {https://doi.org/10.1103/PhysRevLett.109.095703}
  {\bibfield  {journal} {\bibinfo  {journal} {Phys. Rev. Lett.}\ }\textbf
  {\bibinfo {volume} {109}},\ \bibinfo {pages} {095703} (\bibinfo {year}
  {2012})}\BibitemShut {NoStop}%
\bibitem [{\citenamefont {Bertrand}\ \emph {et~al.}(2016)\citenamefont
  {Bertrand}, \citenamefont {Behringer}, \citenamefont {Chakraborty},
  \citenamefont {O'Hern},\ and\ \citenamefont {Shattuck}}]{Bertrand2016_pre}%
  \BibitemOpen
  \bibfield  {author} {\bibinfo {author} {\bibfnamefont {T.}~\bibnamefont
  {Bertrand}}, \bibinfo {author} {\bibfnamefont {R.~P.}\ \bibnamefont
  {Behringer}}, \bibinfo {author} {\bibfnamefont {B.}~\bibnamefont
  {Chakraborty}}, \bibinfo {author} {\bibfnamefont {C.~S.}\ \bibnamefont
  {O'Hern}},\ and\ \bibinfo {author} {\bibfnamefont {M.~D.}\ \bibnamefont
  {Shattuck}},\ }\bibfield  {title} {\emph {\bibinfo {title} {Protocol
  dependence of the jamming transition}},\ }\href
  {https://doi.org/10.1103/PhysRevE.93.012901} {\bibfield  {journal} {\bibinfo
  {journal} {Physical Review E}\ }\textbf {\bibinfo {volume} {93}},\ \bibinfo
  {pages} {012901} (\bibinfo {year} {2016})}\BibitemShut {NoStop}%
\bibitem [{\citenamefont {Baity-Jesi}\ \emph {et~al.}(2017)\citenamefont
  {Baity-Jesi}, \citenamefont {Goodrich}, \citenamefont {Liu}, \citenamefont
  {Nagel},\ and\ \citenamefont {Sethna}}]{Baity2017_jps}%
  \BibitemOpen
  \bibfield  {author} {\bibinfo {author} {\bibfnamefont {M.}~\bibnamefont
  {Baity-Jesi}}, \bibinfo {author} {\bibfnamefont {C.~P.}\ \bibnamefont
  {Goodrich}}, \bibinfo {author} {\bibfnamefont {A.~J.}\ \bibnamefont {Liu}},
  \bibinfo {author} {\bibfnamefont {S.~R.}\ \bibnamefont {Nagel}},\ and\
  \bibinfo {author} {\bibfnamefont {J.~P.}\ \bibnamefont {Sethna}},\ }\bibfield
   {title} {\emph {\bibinfo {title} {Emergent so(3) symmetry of the
  frictionless shear jamming transition}},\ }\href
  {https://doi.org/10.1007/s10955-016-1703-9} {\bibfield  {journal} {\bibinfo
  {journal} {Journal of Statistical Physics}\ }\textbf {\bibinfo {volume}
  {167}},\ \bibinfo {pages} {735} (\bibinfo {year} {2017})}\BibitemShut
  {NoStop}%
\bibitem [{\citenamefont {Zhao}\ \emph
  {et~al.}(2019{\natexlab{a}})\citenamefont {Zhao}, \citenamefont {Bar\'es},
  \citenamefont {Zheng}, \citenamefont {Socolar},\ and\ \citenamefont
  {Behringer}}]{zhao2019_prl}%
  \BibitemOpen
  \bibfield  {author} {\bibinfo {author} {\bibfnamefont {Y.}~\bibnamefont
  {Zhao}}, \bibinfo {author} {\bibfnamefont {J.}~\bibnamefont {Bar\'es}},
  \bibinfo {author} {\bibfnamefont {H.}~\bibnamefont {Zheng}}, \bibinfo
  {author} {\bibfnamefont {J.~E.~S.}\ \bibnamefont {Socolar}},\ and\ \bibinfo
  {author} {\bibfnamefont {R.~P.}\ \bibnamefont {Behringer}},\ }\bibfield
  {title} {\emph {\bibinfo {title} {Shear-jammed, fragile, and steady states in
  homogeneously strained granular materials}},\ }\href
  {https://doi.org/10.1103/PhysRevLett.123.158001} {\bibfield  {journal}
  {\bibinfo  {journal} {Phys. Rev. Lett.}\ }\textbf {\bibinfo {volume} {123}},\
  \bibinfo {pages} {158001} (\bibinfo {year} {2019}{\natexlab{a}})}\BibitemShut
  {NoStop}%
\bibitem [{\citenamefont {Kumar}\ and\ \citenamefont
  {Luding}(2016)}]{Kumar2016_gm}%
  \BibitemOpen
  \bibfield  {author} {\bibinfo {author} {\bibfnamefont {N.}~\bibnamefont
  {Kumar}}\ and\ \bibinfo {author} {\bibfnamefont {S.}~\bibnamefont {Luding}},\
  }\bibfield  {title} {\emph {\bibinfo {title} {Memory of jamming--multiscale
  models for soft and granular matter}},\ }\href
  {https://doi.org/10.1007/s10035-016-0624-2} {\bibfield  {journal} {\bibinfo
  {journal} {Granular Matter}\ }\textbf {\bibinfo {volume} {18}},\ \bibinfo
  {pages} {58} (\bibinfo {year} {2016})}\BibitemShut {NoStop}%
\bibitem [{\citenamefont {Urbani}\ and\ \citenamefont
  {Zamponi}(2017)}]{urbani2017_prl}%
  \BibitemOpen
  \bibfield  {author} {\bibinfo {author} {\bibfnamefont {P.}~\bibnamefont
  {Urbani}}\ and\ \bibinfo {author} {\bibfnamefont {F.}~\bibnamefont
  {Zamponi}},\ }\bibfield  {title} {\emph {\bibinfo {title} {Shear yielding and
  shear jamming of dense hard sphere glasses}},\ }\href
  {https://doi.org/10.1103/PhysRevLett.118.038001} {\bibfield  {journal}
  {\bibinfo  {journal} {Physical Review Letters}\ }\textbf {\bibinfo {volume}
  {118}},\ \bibinfo {pages} {038001} (\bibinfo {year} {2017})}\BibitemShut
  {NoStop}%
\bibitem [{\citenamefont {Jin}\ \emph {et~al.}(2018)\citenamefont {Jin},
  \citenamefont {Urbani}, \citenamefont {Zamponi},\ and\ \citenamefont
  {Yoshino}}]{Jin2018_sa}%
  \BibitemOpen
  \bibfield  {author} {\bibinfo {author} {\bibfnamefont {Y.}~\bibnamefont
  {Jin}}, \bibinfo {author} {\bibfnamefont {P.}~\bibnamefont {Urbani}},
  \bibinfo {author} {\bibfnamefont {F.}~\bibnamefont {Zamponi}},\ and\ \bibinfo
  {author} {\bibfnamefont {H.}~\bibnamefont {Yoshino}},\ }\bibfield  {title}
  {\emph {\bibinfo {title} {A stability-reversibility map unifies elasticity,
  plasticity, yielding, and jamming in hard sphere glasses}},\ }\href
  {https://advances.sciencemag.org/content/4/12/eaat6387} {\bibfield  {journal}
  {\bibinfo  {journal} {Science Advances}\ }\textbf {\bibinfo {volume} {4}}
  (\bibinfo {year} {2018})}\BibitemShut {NoStop}%
\bibitem [{\citenamefont {Otsuki}\ and\ \citenamefont
  {Hayakawa}(2020)}]{otsuki2020_pre}%
  \BibitemOpen
  \bibfield  {author} {\bibinfo {author} {\bibfnamefont {M.}~\bibnamefont
  {Otsuki}}\ and\ \bibinfo {author} {\bibfnamefont {H.}~\bibnamefont
  {Hayakawa}},\ }\bibfield  {title} {\emph {\bibinfo {title} {Shear jamming,
  discontinuous shear thickening, and fragile states in dry granular materials
  under oscillatory shear}},\ }\href
  {https://doi.org/10.1103/PhysRevE.101.032905} {\bibfield  {journal} {\bibinfo
   {journal} {Phys. Rev. E}\ }\textbf {\bibinfo {volume} {101}},\ \bibinfo
  {pages} {032905} (\bibinfo {year} {2020})}\BibitemShut {NoStop}%
\bibitem [{\citenamefont {Jin}\ and\ \citenamefont
  {Yoshino}(2021)}]{Jine2021_pnas}%
  \BibitemOpen
  \bibfield  {author} {\bibinfo {author} {\bibfnamefont {Y.}~\bibnamefont
  {Jin}}\ and\ \bibinfo {author} {\bibfnamefont {H.}~\bibnamefont {Yoshino}},\
  }\bibfield  {title} {\emph {\bibinfo {title} {A jamming plane of sphere
  packings}},\ }\href {https://www.pnas.org/content/118/14/e2021794118}
  {\bibfield  {journal} {\bibinfo  {journal} {Proceedings of the National
  Academy of Sciences}\ }\textbf {\bibinfo {volume} {118}} (\bibinfo {year}
  {2021})}\BibitemShut {NoStop}%
\bibitem [{\citenamefont {Xiong}\ \emph {et~al.}(2019)\citenamefont {Xiong},
  \citenamefont {Wang}, \citenamefont {Clark}, \citenamefont {Bertrand},
  \citenamefont {Ouellette}, \citenamefont {Shattuck},\ and\ \citenamefont
  {O'Hern}}]{Xiong2019_gm}%
  \BibitemOpen
  \bibfield  {author} {\bibinfo {author} {\bibfnamefont {F.}~\bibnamefont
  {Xiong}}, \bibinfo {author} {\bibfnamefont {P.}~\bibnamefont {Wang}},
  \bibinfo {author} {\bibfnamefont {A.~H.}\ \bibnamefont {Clark}}, \bibinfo
  {author} {\bibfnamefont {T.}~\bibnamefont {Bertrand}}, \bibinfo {author}
  {\bibfnamefont {N.~T.}\ \bibnamefont {Ouellette}}, \bibinfo {author}
  {\bibfnamefont {M.~D.}\ \bibnamefont {Shattuck}},\ and\ \bibinfo {author}
  {\bibfnamefont {C.~S.}\ \bibnamefont {O'Hern}},\ }\bibfield  {title} {\emph
  {\bibinfo {title} {Comparison of shear and compression jammed packings of
  frictional disks}},\ }\href {https://doi.org/10.1007/s10035-019-0964-9}
  {\bibfield  {journal} {\bibinfo  {journal} {Granular Matter}\ }\textbf
  {\bibinfo {volume} {21}},\ \bibinfo {pages} {109} (\bibinfo {year}
  {2019})}\BibitemShut {NoStop}%
\bibitem [{\citenamefont {Mari}\ \emph {et~al.}(2014)\citenamefont {Mari},
  \citenamefont {Seto}, \citenamefont {Morris},\ and\ \citenamefont
  {Denn}}]{mari2014_jor}%
  \BibitemOpen
  \bibfield  {author} {\bibinfo {author} {\bibfnamefont {R.}~\bibnamefont
  {Mari}}, \bibinfo {author} {\bibfnamefont {R.}~\bibnamefont {Seto}}, \bibinfo
  {author} {\bibfnamefont {J.~F.}\ \bibnamefont {Morris}},\ and\ \bibinfo
  {author} {\bibfnamefont {M.~M.}\ \bibnamefont {Denn}},\ }\bibfield  {title}
  {\emph {\bibinfo {title} {Shear thickening, frictionless and frictional
  rheologies in non-brownian suspensions}},\ }\href
  {https://doi.org/10.1122/1.4890747} {\bibfield  {journal} {\bibinfo
  {journal} {Journal of Rheology}\ }\textbf {\bibinfo {volume} {58}},\ \bibinfo
  {pages} {1693} (\bibinfo {year} {2014})}\BibitemShut {NoStop}%
\bibitem [{\citenamefont {Wyart}\ and\ \citenamefont
  {Cates}(2014)}]{wyart2014_prl}%
  \BibitemOpen
  \bibfield  {author} {\bibinfo {author} {\bibfnamefont {M.}~\bibnamefont
  {Wyart}}\ and\ \bibinfo {author} {\bibfnamefont {M.~E.}\ \bibnamefont
  {Cates}},\ }\bibfield  {title} {\emph {\bibinfo {title} {Discontinuous shear
  thickening without inertia in dense non-brownian suspensions}},\ }\href
  {https://doi.org/10.1103/PhysRevLett.112.098302} {\bibfield  {journal}
  {\bibinfo  {journal} {Phys. Rev. Lett.}\ }\textbf {\bibinfo {volume} {112}},\
  \bibinfo {pages} {098302} (\bibinfo {year} {2014})}\BibitemShut {NoStop}%
\bibitem [{\citenamefont {Brown}\ and\ \citenamefont
  {Jaeger}(2014)}]{brown2014_RPP}%
  \BibitemOpen
  \bibfield  {author} {\bibinfo {author} {\bibfnamefont {E.}~\bibnamefont
  {Brown}}\ and\ \bibinfo {author} {\bibfnamefont {H.~M.}\ \bibnamefont
  {Jaeger}},\ }\bibfield  {title} {\emph {\bibinfo {title} {Shear thickening in
  concentrated suspensions: phenomenology, mechanisms and relations to
  jamming}},\ }\href {https://doi.org/10.1088/0034-4885/77/4/046602} {\bibfield
   {journal} {\bibinfo  {journal} {Reports on Progress in Physics}\ }\textbf
  {\bibinfo {volume} {77}},\ \bibinfo {pages} {046602} (\bibinfo {year}
  {2014})}\BibitemShut {NoStop}%
\bibitem [{\citenamefont {Han}\ \emph {et~al.}(2018)\citenamefont {Han},
  \citenamefont {Wyart}, \citenamefont {Peters},\ and\ \citenamefont
  {Jaeger}}]{han2018_prf}%
  \BibitemOpen
  \bibfield  {author} {\bibinfo {author} {\bibfnamefont {E.}~\bibnamefont
  {Han}}, \bibinfo {author} {\bibfnamefont {M.}~\bibnamefont {Wyart}}, \bibinfo
  {author} {\bibfnamefont {I.~R.}\ \bibnamefont {Peters}},\ and\ \bibinfo
  {author} {\bibfnamefont {H.~M.}\ \bibnamefont {Jaeger}},\ }\bibfield  {title}
  {\emph {\bibinfo {title} {Shear fronts in shear-thickening suspensions}},\
  }\href {https://doi.org/10.1103/PhysRevFluids.3.073301} {\bibfield  {journal}
  {\bibinfo  {journal} {Phys. Rev. Fluids}\ }\textbf {\bibinfo {volume} {3}},\
  \bibinfo {pages} {073301} (\bibinfo {year} {2018})}\BibitemShut {NoStop}%
\bibitem [{\citenamefont {Blanco}\ \emph {et~al.}(2019)\citenamefont {Blanco},
  \citenamefont {Hodgson}, \citenamefont {Hermes}, \citenamefont {Besseling},
  \citenamefont {Hunter}, \citenamefont {Chaikin}, \citenamefont {Cates},
  \citenamefont {Van~Damme},\ and\ \citenamefont {Poon}}]{blanco2019_pnas}%
  \BibitemOpen
  \bibfield  {author} {\bibinfo {author} {\bibfnamefont {E.}~\bibnamefont
  {Blanco}}, \bibinfo {author} {\bibfnamefont {D.~J.~M.}\ \bibnamefont
  {Hodgson}}, \bibinfo {author} {\bibfnamefont {M.}~\bibnamefont {Hermes}},
  \bibinfo {author} {\bibfnamefont {R.}~\bibnamefont {Besseling}}, \bibinfo
  {author} {\bibfnamefont {G.~L.}\ \bibnamefont {Hunter}}, \bibinfo {author}
  {\bibfnamefont {P.~M.}\ \bibnamefont {Chaikin}}, \bibinfo {author}
  {\bibfnamefont {M.~E.}\ \bibnamefont {Cates}}, \bibinfo {author}
  {\bibfnamefont {I.}~\bibnamefont {Van~Damme}},\ and\ \bibinfo {author}
  {\bibfnamefont {W.~C.~K.}\ \bibnamefont {Poon}},\ }\bibfield  {title} {\emph
  {\bibinfo {title} {Conching chocolate is a prototypical transition from
  frictionally jammed solid to flowable suspension with maximal solid
  content}},\ }\href {https://doi.org/10.1073/pnas.1901858116} {\bibfield
  {journal} {\bibinfo  {journal} {Proceedings of the National Academy of
  Sciences}\ }\textbf {\bibinfo {volume} {116}},\ \bibinfo {pages} {10303}
  (\bibinfo {year} {2019})}\BibitemShut {NoStop}%
\bibitem [{\citenamefont {Morris}(2020)}]{morris2020_arfm}%
  \BibitemOpen
  \bibfield  {author} {\bibinfo {author} {\bibfnamefont {J.~F.}\ \bibnamefont
  {Morris}},\ }\bibfield  {title} {\emph {\bibinfo {title} {Shear thickening of
  concentrated suspensions: Recent developments and relation to other
  phenomena}},\ }\href {https://doi.org/10.1146/annurev-fluid-010816-060128}
  {\bibfield  {journal} {\bibinfo  {journal} {Annual Review of Fluid
  Mechanics}\ }\textbf {\bibinfo {volume} {52}},\ \bibinfo {pages} {121}
  (\bibinfo {year} {2020})}\BibitemShut {NoStop}%
\bibitem [{\citenamefont {Cates}\ \emph {et~al.}(1998)\citenamefont {Cates},
  \citenamefont {Wittmer}, \citenamefont {Bouchaud},\ and\ \citenamefont
  {Claudin}}]{Cates1998_prl}%
  \BibitemOpen
  \bibfield  {author} {\bibinfo {author} {\bibfnamefont {M.~E.}\ \bibnamefont
  {Cates}}, \bibinfo {author} {\bibfnamefont {J.~P.}\ \bibnamefont {Wittmer}},
  \bibinfo {author} {\bibfnamefont {J.-P.}\ \bibnamefont {Bouchaud}},\ and\
  \bibinfo {author} {\bibfnamefont {P.}~\bibnamefont {Claudin}},\ }\bibfield
  {title} {\emph {\bibinfo {title} {Jamming, force chains, and fragile
  matter}},\ }\href {https://doi.org/10.1103/PhysRevLett.81.1841} {\bibfield
  {journal} {\bibinfo  {journal} {Phys. Rev. Lett.}\ }\textbf {\bibinfo
  {volume} {81}},\ \bibinfo {pages} {1841} (\bibinfo {year}
  {1998})}\BibitemShut {NoStop}%
\bibitem [{\citenamefont {Seto}\ \emph {et~al.}(2019)\citenamefont {Seto},
  \citenamefont {Singh}, \citenamefont {Chakraborty}, \citenamefont {Denn},\
  and\ \citenamefont {Morris}}]{seto2019_gm}%
  \BibitemOpen
  \bibfield  {author} {\bibinfo {author} {\bibfnamefont {R.}~\bibnamefont
  {Seto}}, \bibinfo {author} {\bibfnamefont {A.}~\bibnamefont {Singh}},
  \bibinfo {author} {\bibfnamefont {B.}~\bibnamefont {Chakraborty}}, \bibinfo
  {author} {\bibfnamefont {M.~M.}\ \bibnamefont {Denn}},\ and\ \bibinfo
  {author} {\bibfnamefont {J.~F.}\ \bibnamefont {Morris}},\ }\bibfield  {title}
  {\emph {\bibinfo {title} {Shear jamming and fragility in dense
  suspensions}},\ }\href {https://doi.org/10.1007/s10035-019-0931-5} {\bibfield
   {journal} {\bibinfo  {journal} {Granular Matter}\ }\textbf {\bibinfo
  {volume} {21}},\ \bibinfo {pages} {82} (\bibinfo {year} {2019})}\BibitemShut
  {NoStop}%
\bibitem [{\citenamefont {Sarkar}\ \emph {et~al.}(2013)\citenamefont {Sarkar},
  \citenamefont {Bi}, \citenamefont {Zhang}, \citenamefont {Behringer},\ and\
  \citenamefont {Chakraborty}}]{sarkar2013_prl}%
  \BibitemOpen
  \bibfield  {author} {\bibinfo {author} {\bibfnamefont {S.}~\bibnamefont
  {Sarkar}}, \bibinfo {author} {\bibfnamefont {D.}~\bibnamefont {Bi}}, \bibinfo
  {author} {\bibfnamefont {J.}~\bibnamefont {Zhang}}, \bibinfo {author}
  {\bibfnamefont {R.~P.}\ \bibnamefont {Behringer}},\ and\ \bibinfo {author}
  {\bibfnamefont {B.}~\bibnamefont {Chakraborty}},\ }\bibfield  {title} {\emph
  {\bibinfo {title} {Origin of rigidity in dry granular solids}},\ }\href
  {https://doi.org/10.1103/PhysRevLett.111.068301} {\bibfield  {journal}
  {\bibinfo  {journal} {Phys. Rev. Lett.}\ }\textbf {\bibinfo {volume} {111}},\
  \bibinfo {pages} {068301} (\bibinfo {year} {2013})}\BibitemShut {NoStop}%
\bibitem [{\citenamefont {Sarkar}\ \emph {et~al.}(2016)\citenamefont {Sarkar},
  \citenamefont {Bi}, \citenamefont {Zhang}, \citenamefont {Ren}, \citenamefont
  {Behringer},\ and\ \citenamefont {Chakraborty}}]{sarkar2016_pre}%
  \BibitemOpen
  \bibfield  {author} {\bibinfo {author} {\bibfnamefont {S.}~\bibnamefont
  {Sarkar}}, \bibinfo {author} {\bibfnamefont {D.}~\bibnamefont {Bi}}, \bibinfo
  {author} {\bibfnamefont {J.}~\bibnamefont {Zhang}}, \bibinfo {author}
  {\bibfnamefont {J.}~\bibnamefont {Ren}}, \bibinfo {author} {\bibfnamefont
  {R.~P.}\ \bibnamefont {Behringer}},\ and\ \bibinfo {author} {\bibfnamefont
  {B.}~\bibnamefont {Chakraborty}},\ }\bibfield  {title} {\emph {\bibinfo
  {title} {Shear-induced rigidity of frictional particles: Analysis of emergent
  order in stress space}},\ }\href {https://doi.org/10.1103/PhysRevE.93.042901}
  {\bibfield  {journal} {\bibinfo  {journal} {Phys. Rev. E}\ }\textbf {\bibinfo
  {volume} {93}},\ \bibinfo {pages} {042901} (\bibinfo {year}
  {2016})}\BibitemShut {NoStop}%
\bibitem [{\citenamefont {Wang}\ \emph {et~al.}(2018)\citenamefont {Wang},
  \citenamefont {Ren}, \citenamefont {Dijksman}, \citenamefont {Zheng},\ and\
  \citenamefont {Behringer}}]{dong2018_prl}%
  \BibitemOpen
  \bibfield  {author} {\bibinfo {author} {\bibfnamefont {D.}~\bibnamefont
  {Wang}}, \bibinfo {author} {\bibfnamefont {J.}~\bibnamefont {Ren}}, \bibinfo
  {author} {\bibfnamefont {J.~A.}\ \bibnamefont {Dijksman}}, \bibinfo {author}
  {\bibfnamefont {H.}~\bibnamefont {Zheng}},\ and\ \bibinfo {author}
  {\bibfnamefont {R.~P.}\ \bibnamefont {Behringer}},\ }\bibfield  {title}
  {\emph {\bibinfo {title} {Microscopic origins of shear jamming for 2d
  frictional grains}},\ }\href {https://doi.org/10.1103/PhysRevLett.120.208004}
  {\bibfield  {journal} {\bibinfo  {journal} {Phys. Rev. Lett.}\ }\textbf
  {\bibinfo {volume} {120}},\ \bibinfo {pages} {208004} (\bibinfo {year}
  {2018})}\BibitemShut {NoStop}%
\bibitem [{\citenamefont {Royer}\ and\ \citenamefont
  {Chaikin}(2015)}]{Royer2015_pnas}%
  \BibitemOpen
  \bibfield  {author} {\bibinfo {author} {\bibfnamefont {J.~R.}\ \bibnamefont
  {Royer}}\ and\ \bibinfo {author} {\bibfnamefont {P.~M.}\ \bibnamefont
  {Chaikin}},\ }\bibfield  {title} {\emph {\bibinfo {title} {Precisely cyclic
  sand: Self-organization of periodically sheared frictional grains}},\ }\href
  {https://doi.org/10.1073/pnas.1413468112} {\bibfield  {journal} {\bibinfo
  {journal} {Proceedings of the National Academy of Sciences}\ }\textbf
  {\bibinfo {volume} {112}},\ \bibinfo {pages} {49} (\bibinfo {year}
  {2015})}\BibitemShut {NoStop}%
\bibitem [{\citenamefont {Otsuki}\ and\ \citenamefont
  {Hayakawa}(2021)}]{Otsuki2021_epje}%
  \BibitemOpen
  \bibfield  {author} {\bibinfo {author} {\bibfnamefont {M.}~\bibnamefont
  {Otsuki}}\ and\ \bibinfo {author} {\bibfnamefont {H.}~\bibnamefont
  {Hayakawa}},\ }\bibfield  {title} {\emph {\bibinfo {title} {Shear modulus and
  reversible particle trajectories of frictional granular materials under
  oscillatory shear}},\ }\href
  {https://doi.org/10.1140/epje/s10189-021-00075-0} {\bibfield  {journal}
  {\bibinfo  {journal} {The European Physical Journal E}\ }\textbf {\bibinfo
  {volume} {44}},\ \bibinfo {pages} {70} (\bibinfo {year} {2021})}\BibitemShut
  {NoStop}%
\bibitem [{\citenamefont {Keim}\ and\ \citenamefont
  {Arratia}(2014)}]{keim2014_prl}%
  \BibitemOpen
  \bibfield  {author} {\bibinfo {author} {\bibfnamefont {N.~C.}\ \bibnamefont
  {Keim}}\ and\ \bibinfo {author} {\bibfnamefont {P.~E.}\ \bibnamefont
  {Arratia}},\ }\bibfield  {title} {\emph {\bibinfo {title} {Mechanical and
  microscopic properties of the reversible plastic regime in a 2d jammed
  material}},\ }\href {https://doi.org/10.1103/PhysRevLett.112.028302}
  {\bibfield  {journal} {\bibinfo  {journal} {Phys. Rev. Lett.}\ }\textbf
  {\bibinfo {volume} {112}},\ \bibinfo {pages} {028302} (\bibinfo {year}
  {2014})}\BibitemShut {NoStop}%
\bibitem [{\citenamefont {Galloway}\ \emph {et~al.}(2022)\citenamefont
  {Galloway}, \citenamefont {Teich}, \citenamefont {Ma}, \citenamefont
  {Kammer}, \citenamefont {Graham}, \citenamefont {Keim}, \citenamefont
  {Reina}, \citenamefont {Jerolmack}, \citenamefont {Yodh},\ and\ \citenamefont
  {Arratia}}]{Galloway2022_natphy}%
  \BibitemOpen
  \bibfield  {author} {\bibinfo {author} {\bibfnamefont {K.~L.}\ \bibnamefont
  {Galloway}}, \bibinfo {author} {\bibfnamefont {E.~G.}\ \bibnamefont {Teich}},
  \bibinfo {author} {\bibfnamefont {X.~G.}\ \bibnamefont {Ma}}, \bibinfo
  {author} {\bibfnamefont {C.}~\bibnamefont {Kammer}}, \bibinfo {author}
  {\bibfnamefont {I.~R.}\ \bibnamefont {Graham}}, \bibinfo {author}
  {\bibfnamefont {N.~C.}\ \bibnamefont {Keim}}, \bibinfo {author}
  {\bibfnamefont {C.}~\bibnamefont {Reina}}, \bibinfo {author} {\bibfnamefont
  {D.~J.}\ \bibnamefont {Jerolmack}}, \bibinfo {author} {\bibfnamefont {A.~G.}\
  \bibnamefont {Yodh}},\ and\ \bibinfo {author} {\bibfnamefont {P.~E.}\
  \bibnamefont {Arratia}},\ }\bibfield  {title} {\emph {\bibinfo {title}
  {Relationships between structure, memory and flow in sheared disordered
  materials}},\ }\bibfield  {journal} {\bibinfo  {journal} {Nature Physics}\
  }\href {https://doi.org/10.1038/s41567-022-01536-9}
  {10.1038/s41567-022-01536-9} (\bibinfo {year} {2022})\BibitemShut {NoStop}%
\bibitem [{\citenamefont {Keim}\ and\ \citenamefont
  {Medina}(2021)}]{keim2021_arxiv}%
  \BibitemOpen
  \bibfield  {author} {\bibinfo {author} {\bibfnamefont {N.~C.}\ \bibnamefont
  {Keim}}\ and\ \bibinfo {author} {\bibfnamefont {D.}~\bibnamefont {Medina}},\
  }\bibfield  {title} {\emph {\bibinfo {title} {Mechanical annealing and
  memories in a disordered solid}},\ }\href@noop {} {\bibfield  {journal}
  {\bibinfo  {journal} {arXiv preprint arXiv:2112.07008}\ } (\bibinfo {year}
  {2021})}\BibitemShut {NoStop}%
\bibitem [{\citenamefont {Majmudar}\ and\ \citenamefont
  {Behringer}(2005)}]{majmudar2005_nature}%
  \BibitemOpen
  \bibfield  {author} {\bibinfo {author} {\bibfnamefont {T.~S.}\ \bibnamefont
  {Majmudar}}\ and\ \bibinfo {author} {\bibfnamefont {R.~P.}\ \bibnamefont
  {Behringer}},\ }\bibfield  {title} {\emph {\bibinfo {title} {Contact force
  measurements and stress-induced anisotropy in granular materials}},\ }\href
  {https://doi.org/10.1038/nature03805} {\bibfield  {journal} {\bibinfo
  {journal} {Nature}\ }\textbf {\bibinfo {volume} {435}},\ \bibinfo {pages}
  {1079} (\bibinfo {year} {2005})}\BibitemShut {NoStop}%
\bibitem [{\citenamefont {Daniels}\ \emph {et~al.}(2017)\citenamefont
  {Daniels}, \citenamefont {Kollmer},\ and\ \citenamefont
  {Puckett}}]{Daniels2017_rsi}%
  \BibitemOpen
  \bibfield  {author} {\bibinfo {author} {\bibfnamefont {K.~E.}\ \bibnamefont
  {Daniels}}, \bibinfo {author} {\bibfnamefont {J.~E.}\ \bibnamefont
  {Kollmer}},\ and\ \bibinfo {author} {\bibfnamefont {J.~G.}\ \bibnamefont
  {Puckett}},\ }\bibfield  {title} {\emph {\bibinfo {title} {Photoelastic force
  measurements in granular materials}},\ }\href
  {https://doi.org/10.1063/1.4983049} {\bibfield  {journal} {\bibinfo
  {journal} {Review of Scientific Instruments}\ }\textbf {\bibinfo {volume}
  {88}},\ \bibinfo {pages} {051808} (\bibinfo {year} {2017})}\BibitemShut
  {NoStop}%
\bibitem [{\citenamefont {Abed~Zadeh}\ \emph {et~al.}(2019)\citenamefont
  {Abed~Zadeh}, \citenamefont {Bar{\'e}s}, \citenamefont {Brzinski},
  \citenamefont {Daniels}, \citenamefont {Dijksman}, \citenamefont {Docquier},
  \citenamefont {Everitt}, \citenamefont {Kollmer}, \citenamefont {Lantsoght},
  \citenamefont {Wang}, \citenamefont {Workamp}, \citenamefont {Zhao},\ and\
  \citenamefont {Zheng}}]{zadeh2019_gm}%
  \BibitemOpen
  \bibfield  {author} {\bibinfo {author} {\bibfnamefont {A.}~\bibnamefont
  {Abed~Zadeh}}, \bibinfo {author} {\bibfnamefont {J.}~\bibnamefont
  {Bar{\'e}s}}, \bibinfo {author} {\bibfnamefont {T.~A.}\ \bibnamefont
  {Brzinski}}, \bibinfo {author} {\bibfnamefont {K.~E.}\ \bibnamefont
  {Daniels}}, \bibinfo {author} {\bibfnamefont {J.}~\bibnamefont {Dijksman}},
  \bibinfo {author} {\bibfnamefont {N.}~\bibnamefont {Docquier}}, \bibinfo
  {author} {\bibfnamefont {H.~O.}\ \bibnamefont {Everitt}}, \bibinfo {author}
  {\bibfnamefont {J.~E.}\ \bibnamefont {Kollmer}}, \bibinfo {author}
  {\bibfnamefont {O.}~\bibnamefont {Lantsoght}}, \bibinfo {author}
  {\bibfnamefont {D.}~\bibnamefont {Wang}}, \bibinfo {author} {\bibfnamefont
  {M.}~\bibnamefont {Workamp}}, \bibinfo {author} {\bibfnamefont
  {Y.}~\bibnamefont {Zhao}},\ and\ \bibinfo {author} {\bibfnamefont
  {H.}~\bibnamefont {Zheng}},\ }\bibfield  {title} {\emph {\bibinfo {title}
  {Enlightening force chains: a review of photoelasticimetry in granular
  matter}},\ }\href {https://doi.org/10.1007/s10035-019-0942-2} {\bibfield
  {journal} {\bibinfo  {journal} {Granular Matter}\ }\textbf {\bibinfo {volume}
  {21}},\ \bibinfo {pages} {83} (\bibinfo {year} {2019})}\BibitemShut {NoStop}%
\bibitem [{\citenamefont {Andreotti}\ \emph {et~al.}(2013)\citenamefont
  {Andreotti}, \citenamefont {Forterre},\ and\ \citenamefont
  {Pouliquen}}]{andreotti_forterre_pouliquen_2013}%
  \BibitemOpen
  \bibfield  {author} {\bibinfo {author} {\bibfnamefont {B.}~\bibnamefont
  {Andreotti}}, \bibinfo {author} {\bibfnamefont {Y.}~\bibnamefont
  {Forterre}},\ and\ \bibinfo {author} {\bibfnamefont {O.}~\bibnamefont
  {Pouliquen}},\ }\href {https://doi.org/10.1017/CBO9781139541008} {\emph
  {\bibinfo {title} {Granular Media: Between Fluid and Solid}}}\ (\bibinfo
  {publisher} {Cambridge University Press},\ \bibinfo {year}
  {2013})\BibitemShut {NoStop}%
\bibitem [{\citenamefont {Ren}(2013)}]{ren2013_thesis}%
  \BibitemOpen
  \bibfield  {author} {\bibinfo {author} {\bibfnamefont {J.}~\bibnamefont
  {Ren}},\ }\emph {\bibinfo {title} {Nonlinear dynamics and network properties
  in granular materials under shear}},\ \href@noop {} {Ph.D. thesis},\ \bibinfo
   {school} {Duke University} (\bibinfo {year} {2013})\BibitemShut {NoStop}%
\bibitem [{\citenamefont {Ren}\ \emph {et~al.}(2013)\citenamefont {Ren},
  \citenamefont {Dijksman},\ and\ \citenamefont {Behringer}}]{Ren2013_prl}%
  \BibitemOpen
  \bibfield  {author} {\bibinfo {author} {\bibfnamefont {J.}~\bibnamefont
  {Ren}}, \bibinfo {author} {\bibfnamefont {J.~A.}\ \bibnamefont {Dijksman}},\
  and\ \bibinfo {author} {\bibfnamefont {R.~P.}\ \bibnamefont {Behringer}},\
  }\bibfield  {title} {\emph {\bibinfo {title} {Reynolds pressure and
  relaxation in a sheared granular system}},\ }\href
  {https://doi.org/10.1103/PhysRevLett.110.018302} {\bibfield  {journal}
  {\bibinfo  {journal} {Physical Review Letters}\ }\textbf {\bibinfo {volume}
  {110}},\ \bibinfo {pages} {018302} (\bibinfo {year} {2013})}\BibitemShut
  {NoStop}%
\bibitem [{\citenamefont {Wang}(2018)}]{wang2018_phd}%
  \BibitemOpen
  \bibfield  {author} {\bibinfo {author} {\bibfnamefont {D.}~\bibnamefont
  {Wang}},\ }\emph {\bibinfo {title} {Response of Granular Materials to Shear:
  Origins of Shear Jamming, Particle Dynamics, and Effects of Particle
  Properties}},\ \href@noop {} {Ph.D. thesis},\ \bibinfo  {school} {Duke
  University} (\bibinfo {year} {2018})\BibitemShut {NoStop}%
\bibitem [{\citenamefont {Sarkar}\ and\ \citenamefont
  {Chakraborty}(2015)}]{sarkar2015_pre}%
  \BibitemOpen
  \bibfield  {author} {\bibinfo {author} {\bibfnamefont {S.}~\bibnamefont
  {Sarkar}}\ and\ \bibinfo {author} {\bibfnamefont {B.}~\bibnamefont
  {Chakraborty}},\ }\bibfield  {title} {\emph {\bibinfo {title} {Shear-induced
  rigidity in athermal materials: A unified statistical framework}},\ }\href
  {https://doi.org/10.1103/PhysRevE.91.042201} {\bibfield  {journal} {\bibinfo
  {journal} {Phys. Rev. E}\ }\textbf {\bibinfo {volume} {91}},\ \bibinfo
  {pages} {042201} (\bibinfo {year} {2015})}\BibitemShut {NoStop}%
\bibitem [{Note1()}]{Note1}%
  \BibitemOpen
  \bibinfo {note} {See Supplementary Material for videos showing (1) the
  strobed states under cyclic shear in two cases where an ultra-stable state is
  formed, (2) the strobed states under cyclic shear in a case of relaxation to
  an unjammed state, (3) the stress relaxation over time for a shear-jammed
  state when the shear is suddenly stopped, {\protect \color {black} and (4)
  the evolution of force network during an initial shear.}}\BibitemShut {Stop}%
\bibitem [{\citenamefont {Peng}\ \emph {et~al.}(2007)\citenamefont {Peng},
  \citenamefont {Balijepalli}, \citenamefont {Gupta},\ and\ \citenamefont
  {LeBrun}}]{peng2007_JCISE}%
  \BibitemOpen
  \bibfield  {author} {\bibinfo {author} {\bibfnamefont {T.}~\bibnamefont
  {Peng}}, \bibinfo {author} {\bibfnamefont {A.}~\bibnamefont {Balijepalli}},
  \bibinfo {author} {\bibfnamefont {S.~K.}\ \bibnamefont {Gupta}},\ and\
  \bibinfo {author} {\bibfnamefont {T.}~\bibnamefont {LeBrun}},\ }\bibfield
  {title} {\emph {\bibinfo {title} {{Algorithms for On-Line Monitoring of Micro
  Spheres in an Optical Tweezers-Based Assembly Cell}}},\ }\href
  {https://doi.org/10.1115/1.2795306} {\bibfield  {journal} {\bibinfo
  {journal} {Journal of Computing and Information Science in Engineering}\
  }\textbf {\bibinfo {volume} {7}},\ \bibinfo {pages} {330} (\bibinfo {year}
  {2007})}\BibitemShut {NoStop}%
\bibitem [{\citenamefont {Howell}\ \emph {et~al.}(1999)\citenamefont {Howell},
  \citenamefont {Behringer},\ and\ \citenamefont {Veje}}]{Howell1999_prl}%
  \BibitemOpen
  \bibfield  {author} {\bibinfo {author} {\bibfnamefont {D.}~\bibnamefont
  {Howell}}, \bibinfo {author} {\bibfnamefont {R.~P.}\ \bibnamefont
  {Behringer}},\ and\ \bibinfo {author} {\bibfnamefont {C.}~\bibnamefont
  {Veje}},\ }\bibfield  {title} {\emph {\bibinfo {title} {Stress fluctuations
  in a 2d granular couette experiment: A continuous transition}},\ }\href
  {https://doi.org/10.1103/PhysRevLett.82.5241} {\bibfield  {journal} {\bibinfo
   {journal} {Phys. Rev. Lett.}\ }\textbf {\bibinfo {volume} {82}},\ \bibinfo
  {pages} {5241} (\bibinfo {year} {1999})}\BibitemShut {NoStop}%
\bibitem [{\citenamefont {Zhao}\ \emph
  {et~al.}(2019{\natexlab{b}})\citenamefont {Zhao}, \citenamefont {Zheng},
  \citenamefont {Wang}, \citenamefont {Wang},\ and\ \citenamefont
  {Behringer}}]{zhao2019_njp}%
  \BibitemOpen
  \bibfield  {author} {\bibinfo {author} {\bibfnamefont {Y.}~\bibnamefont
  {Zhao}}, \bibinfo {author} {\bibfnamefont {H.}~\bibnamefont {Zheng}},
  \bibinfo {author} {\bibfnamefont {D.}~\bibnamefont {Wang}}, \bibinfo {author}
  {\bibfnamefont {M.}~\bibnamefont {Wang}},\ and\ \bibinfo {author}
  {\bibfnamefont {R.~P.}\ \bibnamefont {Behringer}},\ }\bibfield  {title}
  {\emph {\bibinfo {title} {Particle scale force sensor based on intensity
  gradient method in granular photoelastic experiments}},\ }\href
  {https://doi.org/10.1088/1367-2630/ab05e7} {\bibfield  {journal} {\bibinfo
  {journal} {New Journal of Physics}\ }\textbf {\bibinfo {volume} {21}},\
  \bibinfo {pages} {023009} (\bibinfo {year} {2019}{\natexlab{b}})}\BibitemShut
  {NoStop}%
\bibitem [{\citenamefont {Christoffersen}\ \emph {et~al.}(1981)\citenamefont
  {Christoffersen}, \citenamefont {Mehrabadi},\ and\ \citenamefont
  {Nemat-Nasser}}]{christoffersen1981_jam}%
  \BibitemOpen
  \bibfield  {author} {\bibinfo {author} {\bibfnamefont {J.}~\bibnamefont
  {Christoffersen}}, \bibinfo {author} {\bibfnamefont {M.~M.}\ \bibnamefont
  {Mehrabadi}},\ and\ \bibinfo {author} {\bibfnamefont {S.}~\bibnamefont
  {Nemat-Nasser}},\ }\bibfield  {title} {\emph {\bibinfo {title} {{A
  Micromechanical Description of Granular Material Behavior}}},\ }\href
  {https://doi.org/10.1115/1.3157619} {\bibfield  {journal} {\bibinfo
  {journal} {Journal of Applied Mechanics}\ }\textbf {\bibinfo {volume} {48}},\
  \bibinfo {pages} {339} (\bibinfo {year} {1981})}\BibitemShut {NoStop}%
\bibitem [{\citenamefont {Radjai}\ \emph {et~al.}(1998)\citenamefont {Radjai},
  \citenamefont {Wolf}, \citenamefont {Jean},\ and\ \citenamefont
  {Moreau}}]{radjai1998_prl}%
  \BibitemOpen
  \bibfield  {author} {\bibinfo {author} {\bibfnamefont {F.}~\bibnamefont
  {Radjai}}, \bibinfo {author} {\bibfnamefont {D.~E.}\ \bibnamefont {Wolf}},
  \bibinfo {author} {\bibfnamefont {M.}~\bibnamefont {Jean}},\ and\ \bibinfo
  {author} {\bibfnamefont {J.-J.}\ \bibnamefont {Moreau}},\ }\bibfield  {title}
  {\emph {\bibinfo {title} {Bimodal character of stress transmission in
  granular packings}},\ }\href {https://doi.org/10.1103/PhysRevLett.80.61}
  {\bibfield  {journal} {\bibinfo  {journal} {Phys. Rev. Lett.}\ }\textbf
  {\bibinfo {volume} {80}},\ \bibinfo {pages} {61} (\bibinfo {year}
  {1998})}\BibitemShut {NoStop}%
\bibitem [{\citenamefont {Kawasaki}\ and\ \citenamefont
  {Berthier}(2016)}]{kawasaki2016_pre}%
  \BibitemOpen
  \bibfield  {author} {\bibinfo {author} {\bibfnamefont {T.}~\bibnamefont
  {Kawasaki}}\ and\ \bibinfo {author} {\bibfnamefont {L.}~\bibnamefont
  {Berthier}},\ }\bibfield  {title} {\emph {\bibinfo {title} {Macroscopic
  yielding in jammed solids is accompanied by a nonequilibrium first-order
  transition in particle trajectories}},\ }\href
  {https://doi.org/10.1103/PhysRevE.94.022615} {\bibfield  {journal} {\bibinfo
  {journal} {Phys. Rev. E}\ }\textbf {\bibinfo {volume} {94}},\ \bibinfo
  {pages} {022615} (\bibinfo {year} {2016})}\BibitemShut {NoStop}%
\bibitem [{\citenamefont {Kou}\ \emph {et~al.}(2017)\citenamefont {Kou},
  \citenamefont {Cao}, \citenamefont {Li}, \citenamefont {Xia}, \citenamefont
  {Li}, \citenamefont {Dong}, \citenamefont {Zhang}, \citenamefont {Zhang},
  \citenamefont {Kob},\ and\ \citenamefont {Wang}}]{kou2017_nature}%
  \BibitemOpen
  \bibfield  {author} {\bibinfo {author} {\bibfnamefont {B.}~\bibnamefont
  {Kou}}, \bibinfo {author} {\bibfnamefont {Y.}~\bibnamefont {Cao}}, \bibinfo
  {author} {\bibfnamefont {J.}~\bibnamefont {Li}}, \bibinfo {author}
  {\bibfnamefont {C.}~\bibnamefont {Xia}}, \bibinfo {author} {\bibfnamefont
  {Z.}~\bibnamefont {Li}}, \bibinfo {author} {\bibfnamefont {H.}~\bibnamefont
  {Dong}}, \bibinfo {author} {\bibfnamefont {A.}~\bibnamefont {Zhang}},
  \bibinfo {author} {\bibfnamefont {J.}~\bibnamefont {Zhang}}, \bibinfo
  {author} {\bibfnamefont {W.}~\bibnamefont {Kob}},\ and\ \bibinfo {author}
  {\bibfnamefont {Y.}~\bibnamefont {Wang}},\ }\bibfield  {title} {\emph
  {\bibinfo {title} {Granular materials flow like complex fluids}},\ }\href
  {https://doi.org/10.1038/nature24062} {\bibfield  {journal} {\bibinfo
  {journal} {Nature}\ }\textbf {\bibinfo {volume} {551}},\ \bibinfo {pages}
  {360} (\bibinfo {year} {2017})}\BibitemShut {NoStop}%
\bibitem [{\citenamefont {Sun}\ \emph {et~al.}(2020)\citenamefont {Sun},
  \citenamefont {Kob}, \citenamefont {Blumenfeld}, \citenamefont {Tong},
  \citenamefont {Wang},\ and\ \citenamefont {Zhang}}]{sun2020_prl}%
  \BibitemOpen
  \bibfield  {author} {\bibinfo {author} {\bibfnamefont {X.}~\bibnamefont
  {Sun}}, \bibinfo {author} {\bibfnamefont {W.}~\bibnamefont {Kob}}, \bibinfo
  {author} {\bibfnamefont {R.}~\bibnamefont {Blumenfeld}}, \bibinfo {author}
  {\bibfnamefont {H.}~\bibnamefont {Tong}}, \bibinfo {author} {\bibfnamefont
  {Y.}~\bibnamefont {Wang}},\ and\ \bibinfo {author} {\bibfnamefont
  {J.}~\bibnamefont {Zhang}},\ }\bibfield  {title} {\emph {\bibinfo {title}
  {Friction-controlled entropy-stability competition in granular systems}},\
  }\href {https://doi.org/10.1103/PhysRevLett.125.268005} {\bibfield  {journal}
  {\bibinfo  {journal} {Phys. Rev. Lett.}\ }\textbf {\bibinfo {volume} {125}},\
  \bibinfo {pages} {268005} (\bibinfo {year} {2020})}\BibitemShut {NoStop}%
\bibitem [{\citenamefont {Leishangthem}\ \emph {et~al.}(2017)\citenamefont
  {Leishangthem}, \citenamefont {Parmar},\ and\ \citenamefont
  {Sastry}}]{leishangthem2017_natcom}%
  \BibitemOpen
  \bibfield  {author} {\bibinfo {author} {\bibfnamefont {P.}~\bibnamefont
  {Leishangthem}}, \bibinfo {author} {\bibfnamefont {A.~D.~S.}\ \bibnamefont
  {Parmar}},\ and\ \bibinfo {author} {\bibfnamefont {S.}~\bibnamefont
  {Sastry}},\ }\bibfield  {title} {\emph {\bibinfo {title} {The yielding
  transition in amorphous solids under oscillatory shear deformation}},\ }\href
  {https://doi.org/10.1038/ncomms14653} {\bibfield  {journal} {\bibinfo
  {journal} {Nature Communications}\ }\textbf {\bibinfo {volume} {8}},\
  \bibinfo {pages} {14653} (\bibinfo {year} {2017})}\BibitemShut {NoStop}%
\bibitem [{\citenamefont {Lundberg}\ \emph {et~al.}(2008)\citenamefont
  {Lundberg}, \citenamefont {Krishan}, \citenamefont {Xu}, \citenamefont
  {O'Hern},\ and\ \citenamefont {Dennin}}]{Lundberg2008_pre}%
  \BibitemOpen
  \bibfield  {author} {\bibinfo {author} {\bibfnamefont {M.}~\bibnamefont
  {Lundberg}}, \bibinfo {author} {\bibfnamefont {K.}~\bibnamefont {Krishan}},
  \bibinfo {author} {\bibfnamefont {N.}~\bibnamefont {Xu}}, \bibinfo {author}
  {\bibfnamefont {C.~S.}\ \bibnamefont {O'Hern}},\ and\ \bibinfo {author}
  {\bibfnamefont {M.}~\bibnamefont {Dennin}},\ }\bibfield  {title} {\emph
  {\bibinfo {title} {Reversible plastic events in amorphous materials}},\
  }\href {https://doi.org/10.1103/PhysRevE.77.041505} {\bibfield  {journal}
  {\bibinfo  {journal} {Phys. Rev. E}\ }\textbf {\bibinfo {volume} {77}},\
  \bibinfo {pages} {041505} (\bibinfo {year} {2008})}\BibitemShut {NoStop}%
\bibitem [{\citenamefont {Ness}\ and\ \citenamefont
  {Cates}(2020)}]{ness2020_prl}%
  \BibitemOpen
  \bibfield  {author} {\bibinfo {author} {\bibfnamefont {C.}~\bibnamefont
  {Ness}}\ and\ \bibinfo {author} {\bibfnamefont {M.~E.}\ \bibnamefont
  {Cates}},\ }\bibfield  {title} {\emph {\bibinfo {title} {Absorbing-state
  transitions in granular materials close to jamming}},\ }\href
  {https://doi.org/10.1103/PhysRevLett.124.088004} {\bibfield  {journal}
  {\bibinfo  {journal} {Phys. Rev. Lett.}\ }\textbf {\bibinfo {volume} {124}},\
  \bibinfo {pages} {088004} (\bibinfo {year} {2020})}\BibitemShut {NoStop}%
\bibitem [{\citenamefont {Fiocco}\ \emph {et~al.}(2013)\citenamefont {Fiocco},
  \citenamefont {Foffi},\ and\ \citenamefont {Sastry}}]{fiocco2013_pre}%
  \BibitemOpen
  \bibfield  {author} {\bibinfo {author} {\bibfnamefont {D.}~\bibnamefont
  {Fiocco}}, \bibinfo {author} {\bibfnamefont {G.}~\bibnamefont {Foffi}},\ and\
  \bibinfo {author} {\bibfnamefont {S.}~\bibnamefont {Sastry}},\ }\bibfield
  {title} {\emph {\bibinfo {title} {Oscillatory athermal quasistatic
  deformation of a model glass}},\ }\href
  {https://link.aps.org/doi/10.1103/PhysRevE.88.020301} {\bibfield  {journal}
  {\bibinfo  {journal} {Phys. Rev. E}\ }\textbf {\bibinfo {volume} {88}},\
  \bibinfo {pages} {020301} (\bibinfo {year} {2013})}\BibitemShut {NoStop}%
\bibitem [{\citenamefont {Keim}\ \emph {et~al.}(2019)\citenamefont {Keim},
  \citenamefont {Paulsen}, \citenamefont {Zeravcic}, \citenamefont {Sastry},\
  and\ \citenamefont {Nagel}}]{keim2019_rmp}%
  \BibitemOpen
  \bibfield  {author} {\bibinfo {author} {\bibfnamefont {N.~C.}\ \bibnamefont
  {Keim}}, \bibinfo {author} {\bibfnamefont {J.~D.}\ \bibnamefont {Paulsen}},
  \bibinfo {author} {\bibfnamefont {Z.}~\bibnamefont {Zeravcic}}, \bibinfo
  {author} {\bibfnamefont {S.}~\bibnamefont {Sastry}},\ and\ \bibinfo {author}
  {\bibfnamefont {S.~R.}\ \bibnamefont {Nagel}},\ }\bibfield  {title} {\emph
  {\bibinfo {title} {Memory formation in matter}},\ }\href
  {https://doi.org/10.1103/RevModPhys.91.035002} {\bibfield  {journal}
  {\bibinfo  {journal} {Rev. Mod. Phys.}\ }\textbf {\bibinfo {volume} {91}},\
  \bibinfo {pages} {035002} (\bibinfo {year} {2019})}\BibitemShut {NoStop}%
\bibitem [{\citenamefont {Arceri}\ \emph {et~al.}(2021)\citenamefont {Arceri},
  \citenamefont {Corwin},\ and\ \citenamefont {Hagh}}]{arceri2021_pre2}%
  \BibitemOpen
  \bibfield  {author} {\bibinfo {author} {\bibfnamefont {F.}~\bibnamefont
  {Arceri}}, \bibinfo {author} {\bibfnamefont {E.~I.}\ \bibnamefont {Corwin}},\
  and\ \bibinfo {author} {\bibfnamefont {V.~F.}\ \bibnamefont {Hagh}},\
  }\bibfield  {title} {\emph {\bibinfo {title} {Marginal stability in memory
  training of jammed solids}},\ }\href
  {https://doi.org/10.1103/PhysRevE.104.044907} {\bibfield  {journal} {\bibinfo
   {journal} {Phys. Rev. E}\ }\textbf {\bibinfo {volume} {104}},\ \bibinfo
  {pages} {044907} (\bibinfo {year} {2021})}\BibitemShut {NoStop}%
\bibitem [{\citenamefont {Mukherji}\ \emph {et~al.}(2019)\citenamefont
  {Mukherji}, \citenamefont {Kandula}, \citenamefont {Sood},\ and\
  \citenamefont {Ganapathy}}]{Mukherji2019_prl}%
  \BibitemOpen
  \bibfield  {author} {\bibinfo {author} {\bibfnamefont {S.}~\bibnamefont
  {Mukherji}}, \bibinfo {author} {\bibfnamefont {N.}~\bibnamefont {Kandula}},
  \bibinfo {author} {\bibfnamefont {A.~K.}\ \bibnamefont {Sood}},\ and\
  \bibinfo {author} {\bibfnamefont {R.}~\bibnamefont {Ganapathy}},\ }\bibfield
  {title} {\emph {\bibinfo {title} {Strength of mechanical memories is maximal
  at the yield point of a soft glass}},\ }\href
  {https://doi.org/10.1103/PhysRevLett.122.158001} {\bibfield  {journal}
  {\bibinfo  {journal} {Phys. Rev. Lett.}\ }\textbf {\bibinfo {volume} {122}},\
  \bibinfo {pages} {158001} (\bibinfo {year} {2019})}\BibitemShut {NoStop}%
\bibitem [{\citenamefont {Das}\ \emph {et~al.}(2018)\citenamefont {Das},
  \citenamefont {Parmar},\ and\ \citenamefont {Sastry}}]{das2018_arxiv}%
  \BibitemOpen
  \bibfield  {author} {\bibinfo {author} {\bibfnamefont {P.}~\bibnamefont
  {Das}}, \bibinfo {author} {\bibfnamefont {A.~D.}\ \bibnamefont {Parmar}},\
  and\ \bibinfo {author} {\bibfnamefont {S.}~\bibnamefont {Sastry}},\
  }\bibfield  {title} {\emph {\bibinfo {title} {Annealing glasses by cyclic
  shear deformation}},\ }\href@noop {} {\bibfield  {journal} {\bibinfo
  {journal} {arXiv preprint arXiv:1805.12476}\ } (\bibinfo {year}
  {2018})}\BibitemShut {NoStop}%
\bibitem [{\citenamefont {Yeh}\ \emph {et~al.}(2020)\citenamefont {Yeh},
  \citenamefont {Ozawa}, \citenamefont {Miyazaki}, \citenamefont {Kawasaki},\
  and\ \citenamefont {Berthier}}]{yeh2020_prl}%
  \BibitemOpen
  \bibfield  {author} {\bibinfo {author} {\bibfnamefont {W.-T.}\ \bibnamefont
  {Yeh}}, \bibinfo {author} {\bibfnamefont {M.}~\bibnamefont {Ozawa}}, \bibinfo
  {author} {\bibfnamefont {K.}~\bibnamefont {Miyazaki}}, \bibinfo {author}
  {\bibfnamefont {T.}~\bibnamefont {Kawasaki}},\ and\ \bibinfo {author}
  {\bibfnamefont {L.}~\bibnamefont {Berthier}},\ }\bibfield  {title} {\emph
  {\bibinfo {title} {Glass stability changes the nature of yielding under
  oscillatory shear}},\ }\href {https://doi.org/10.1103/PhysRevLett.124.225502}
  {\bibfield  {journal} {\bibinfo  {journal} {Phys. Rev. Lett.}\ }\textbf
  {\bibinfo {volume} {124}},\ \bibinfo {pages} {225502} (\bibinfo {year}
  {2020})}\BibitemShut {NoStop}%
\bibitem [{\citenamefont {Ozawa}\ \emph {et~al.}(2018)\citenamefont {Ozawa},
  \citenamefont {Berthier}, \citenamefont {Biroli}, \citenamefont {Rosso},\
  and\ \citenamefont {Tarjus}}]{ozawa2018_pnas}%
  \BibitemOpen
  \bibfield  {author} {\bibinfo {author} {\bibfnamefont {M.}~\bibnamefont
  {Ozawa}}, \bibinfo {author} {\bibfnamefont {L.}~\bibnamefont {Berthier}},
  \bibinfo {author} {\bibfnamefont {G.}~\bibnamefont {Biroli}}, \bibinfo
  {author} {\bibfnamefont {A.}~\bibnamefont {Rosso}},\ and\ \bibinfo {author}
  {\bibfnamefont {G.}~\bibnamefont {Tarjus}},\ }\bibfield  {title} {\emph
  {\bibinfo {title} {Random critical point separates brittle and ductile
  yielding transitions in amorphous materials}},\ }\href
  {https://doi.org/10.1073/pnas.1806156115} {\bibfield  {journal} {\bibinfo
  {journal} {Proceedings of the National Academy of Sciences}\ }\textbf
  {\bibinfo {volume} {115}},\ \bibinfo {pages} {6656} (\bibinfo {year}
  {2018})}\BibitemShut {NoStop}%
\bibitem [{\citenamefont {Peters}\ \emph {et~al.}(2016)\citenamefont {Peters},
  \citenamefont {Majumdar},\ and\ \citenamefont {Jaeger}}]{Peters2016_nat}%
  \BibitemOpen
  \bibfield  {author} {\bibinfo {author} {\bibfnamefont {I.~R.}\ \bibnamefont
  {Peters}}, \bibinfo {author} {\bibfnamefont {S.}~\bibnamefont {Majumdar}},\
  and\ \bibinfo {author} {\bibfnamefont {H.~M.}\ \bibnamefont {Jaeger}},\
  }\bibfield  {title} {\emph {\bibinfo {title} {Direct observation of dynamic
  shear jamming in dense suspensions}},\ }\href@noop {} {\bibfield  {journal}
  {\bibinfo  {journal} {Nature}\ }\textbf {\bibinfo {volume} {532}},\ \bibinfo
  {pages} {214} (\bibinfo {year} {2016})}\BibitemShut {NoStop}%
\bibitem [{\citenamefont {Han}\ \emph {et~al.}(2019)\citenamefont {Han},
  \citenamefont {James},\ and\ \citenamefont {Jaeger}}]{han2019_prl}%
  \BibitemOpen
  \bibfield  {author} {\bibinfo {author} {\bibfnamefont {E.}~\bibnamefont
  {Han}}, \bibinfo {author} {\bibfnamefont {N.~M.}\ \bibnamefont {James}},\
  and\ \bibinfo {author} {\bibfnamefont {H.~M.}\ \bibnamefont {Jaeger}},\
  }\bibfield  {title} {\emph {\bibinfo {title} {Stress controlled rheology of
  dense suspensions using transient flows}},\ }\href
  {https://doi.org/10.1103/PhysRevLett.123.248002} {\bibfield  {journal}
  {\bibinfo  {journal} {Phys. Rev. Lett.}\ }\textbf {\bibinfo {volume} {123}},\
  \bibinfo {pages} {248002} (\bibinfo {year} {2019})}\BibitemShut {NoStop}%
\bibitem [{\citenamefont {Luding}(2016)}]{luding2016_nature}%
  \BibitemOpen
  \bibfield  {author} {\bibinfo {author} {\bibfnamefont {S.}~\bibnamefont
  {Luding}},\ }\bibfield  {title} {\emph {\bibinfo {title} {So much for the
  jamming point}},\ }\href {https://doi.org/10.1038/nphys3680} {\bibfield
  {journal} {\bibinfo  {journal} {Nature Physics}\ }\textbf {\bibinfo {volume}
  {12}},\ \bibinfo {pages} {531} (\bibinfo {year} {2016})}\BibitemShut
  {NoStop}%
\bibitem [{\citenamefont {Babu}\ \emph {et~al.}(2021)\citenamefont {Babu},
  \citenamefont {Pan}, \citenamefont {Jin}, \citenamefont {Chakraborty},\ and\
  \citenamefont {Sastry}}]{Babu_2021Soft}%
  \BibitemOpen
  \bibfield  {author} {\bibinfo {author} {\bibfnamefont {V.}~\bibnamefont
  {Babu}}, \bibinfo {author} {\bibfnamefont {D.}~\bibnamefont {Pan}}, \bibinfo
  {author} {\bibfnamefont {Y.}~\bibnamefont {Jin}}, \bibinfo {author}
  {\bibfnamefont {B.}~\bibnamefont {Chakraborty}},\ and\ \bibinfo {author}
  {\bibfnamefont {S.}~\bibnamefont {Sastry}},\ }\bibfield  {title} {\emph
  {\bibinfo {title} {Dilatancy{,} shear jamming{,} and a generalized jamming
  phase diagram of frictionless sphere packings}},\ }\href
  {https://doi.org/10.1039/D0SM02186E} {\bibfield  {journal} {\bibinfo
  {journal} {Soft Matter}\ }\textbf {\bibinfo {volume} {17}},\ \bibinfo {pages}
  {3121} (\bibinfo {year} {2021})}\BibitemShut {NoStop}%
\bibitem [{\citenamefont {Das}\ \emph {et~al.}(2020)\citenamefont {Das},
  \citenamefont {Vinutha},\ and\ \citenamefont {Sastry}}]{das2020_pnas}%
  \BibitemOpen
  \bibfield  {author} {\bibinfo {author} {\bibfnamefont {P.}~\bibnamefont
  {Das}}, \bibinfo {author} {\bibfnamefont {H.~A.}\ \bibnamefont {Vinutha}},\
  and\ \bibinfo {author} {\bibfnamefont {S.}~\bibnamefont {Sastry}},\
  }\bibfield  {title} {\emph {\bibinfo {title} {Unified phase diagram of
  reversible{\textendash}irreversible, jamming, and yielding transitions in
  cyclically sheared soft-sphere packings}},\ }\href
  {https://doi.org/10.1073/pnas.1912482117} {\bibfield  {journal} {\bibinfo
  {journal} {Proceedings of the National Academy of Sciences}\ }\textbf
  {\bibinfo {volume} {117}},\ \bibinfo {pages} {10203} (\bibinfo {year}
  {2020})}\BibitemShut {NoStop}%
\bibitem [{\citenamefont {Yang}\ \emph {et~al.}(2021)\citenamefont {Yang},
  \citenamefont {Taiebat}, \citenamefont {Mutabaruka},\ and\ \citenamefont
  {Radja\"{\i}}}]{ming2021_pre}%
  \BibitemOpen
  \bibfield  {author} {\bibinfo {author} {\bibfnamefont {M.}~\bibnamefont
  {Yang}}, \bibinfo {author} {\bibfnamefont {M.}~\bibnamefont {Taiebat}},
  \bibinfo {author} {\bibfnamefont {P.}~\bibnamefont {Mutabaruka}},\ and\
  \bibinfo {author} {\bibfnamefont {F.}~\bibnamefont {Radja\"{\i}}},\
  }\bibfield  {title} {\emph {\bibinfo {title} {Evolution of granular materials
  under isochoric cyclic simple shearing}},\ }\href
  {https://doi.org/10.1103/PhysRevE.103.032904} {\bibfield  {journal} {\bibinfo
   {journal} {Phys. Rev. E}\ }\textbf {\bibinfo {volume} {103}},\ \bibinfo
  {pages} {032904} (\bibinfo {year} {2021})}\BibitemShut {NoStop}%
\bibitem [{\citenamefont {Fardad~Amini}\ \emph {et~al.}(2021)\citenamefont
  {Fardad~Amini}, \citenamefont {Huang},\ and\ \citenamefont
  {Wang}}]{amini2021_GL}%
  \BibitemOpen
  \bibfield  {author} {\bibinfo {author} {\bibfnamefont {P.}~\bibnamefont
  {Fardad~Amini}}, \bibinfo {author} {\bibfnamefont {D.}~\bibnamefont
  {Huang}},\ and\ \bibinfo {author} {\bibfnamefont {G.}~\bibnamefont {Wang}},\
  }\bibfield  {title} {\emph {\bibinfo {title} {Dynamic properties of toyoura
  sand in reliquefaction tests}},\ }\href
  {https://doi.org/10.1680/jgele.20.00099} {\bibfield  {journal} {\bibinfo
  {journal} {G\'{e}otechnique Letters}\ }\textbf {\bibinfo {volume} {11}},\
  \bibinfo {pages} {239} (\bibinfo {year} {2021})}\BibitemShut {NoStop}%
\bibitem [{\citenamefont {Gibaud}\ \emph {et~al.}(2020)\citenamefont {Gibaud},
  \citenamefont {Dag\`es}, \citenamefont {Lidon}, \citenamefont {Jung},
  \citenamefont {Ahour\'e}, \citenamefont {Sztucki}, \citenamefont
  {Poulesquen}, \citenamefont {Hengl}, \citenamefont {Pignon},\ and\
  \citenamefont {Manneville}}]{Gibaud2020_prx}%
  \BibitemOpen
  \bibfield  {author} {\bibinfo {author} {\bibfnamefont {T.}~\bibnamefont
  {Gibaud}}, \bibinfo {author} {\bibfnamefont {N.}~\bibnamefont {Dag\`es}},
  \bibinfo {author} {\bibfnamefont {P.}~\bibnamefont {Lidon}}, \bibinfo
  {author} {\bibfnamefont {G.}~\bibnamefont {Jung}}, \bibinfo {author}
  {\bibfnamefont {L.~C.}\ \bibnamefont {Ahour\'e}}, \bibinfo {author}
  {\bibfnamefont {M.}~\bibnamefont {Sztucki}}, \bibinfo {author} {\bibfnamefont
  {A.}~\bibnamefont {Poulesquen}}, \bibinfo {author} {\bibfnamefont
  {N.}~\bibnamefont {Hengl}}, \bibinfo {author} {\bibfnamefont
  {F.}~\bibnamefont {Pignon}},\ and\ \bibinfo {author} {\bibfnamefont
  {S.}~\bibnamefont {Manneville}},\ }\bibfield  {title} {\emph {\bibinfo
  {title} {Rheoacoustic gels: Tuning mechanical and flow properties of
  colloidal gels with ultrasonic vibrations}},\ }\href
  {https://doi.org/10.1103/PhysRevX.10.011028} {\bibfield  {journal} {\bibinfo
  {journal} {Phys. Rev. X}\ }\textbf {\bibinfo {volume} {10}},\ \bibinfo
  {pages} {011028} (\bibinfo {year} {2020})}\BibitemShut {NoStop}%
\bibitem [{\citenamefont {Dag\`es}\ \emph {et~al.}(2021)\citenamefont
  {Dag\`es}, \citenamefont {Lidon}, \citenamefont {Jung}, \citenamefont
  {Pignon}, \citenamefont {Manneville},\ and\ \citenamefont
  {Gibaud}}]{Dages2021_jor}%
  \BibitemOpen
  \bibfield  {author} {\bibinfo {author} {\bibfnamefont {N.}~\bibnamefont
  {Dag\`es}}, \bibinfo {author} {\bibfnamefont {P.}~\bibnamefont {Lidon}},
  \bibinfo {author} {\bibfnamefont {G.}~\bibnamefont {Jung}}, \bibinfo {author}
  {\bibfnamefont {F.}~\bibnamefont {Pignon}}, \bibinfo {author} {\bibfnamefont
  {S.}~\bibnamefont {Manneville}},\ and\ \bibinfo {author} {\bibfnamefont
  {T.}~\bibnamefont {Gibaud}},\ }\bibfield  {title} {\emph {\bibinfo {title}
  {Mechanics and structure of carbon black gels under high-power ultrasound}},\
  }\href {https://doi.org/10.1122/8.0000187} {\bibfield  {journal} {\bibinfo
  {journal} {Journal of Rheology}\ }\textbf {\bibinfo {volume} {65}},\ \bibinfo
  {pages} {477} (\bibinfo {year} {2021})}\BibitemShut {NoStop}%
\bibitem [{\citenamefont {Lin}\ \emph {et~al.}(2016)\citenamefont {Lin},
  \citenamefont {Ness}, \citenamefont {Cates}, \citenamefont {Sun},\ and\
  \citenamefont {Cohen}}]{lin2016_pnas}%
  \BibitemOpen
  \bibfield  {author} {\bibinfo {author} {\bibfnamefont {N.~Y.}\ \bibnamefont
  {Lin}}, \bibinfo {author} {\bibfnamefont {C.}~\bibnamefont {Ness}}, \bibinfo
  {author} {\bibfnamefont {M.~E.}\ \bibnamefont {Cates}}, \bibinfo {author}
  {\bibfnamefont {J.}~\bibnamefont {Sun}},\ and\ \bibinfo {author}
  {\bibfnamefont {I.}~\bibnamefont {Cohen}},\ }\bibfield  {title} {\emph
  {\bibinfo {title} {Tunable shear thickening in suspensions}},\ }\href
  {https://doi.org/10.1073/pnas.1608348113} {\bibfield  {journal} {\bibinfo
  {journal} {Proceedings of the National Academy of Sciences}\ }\textbf
  {\bibinfo {volume} {113}},\ \bibinfo {pages} {10774} (\bibinfo {year}
  {2016})}\BibitemShut {NoStop}%
\bibitem [{\citenamefont {Ness}\ \emph {et~al.}(2018)\citenamefont {Ness},
  \citenamefont {Mari},\ and\ \citenamefont {Cates}}]{ness2018_sa}%
  \BibitemOpen
  \bibfield  {author} {\bibinfo {author} {\bibfnamefont {C.}~\bibnamefont
  {Ness}}, \bibinfo {author} {\bibfnamefont {R.}~\bibnamefont {Mari}},\ and\
  \bibinfo {author} {\bibfnamefont {M.~E.}\ \bibnamefont {Cates}},\ }\bibfield
  {title} {\emph {\bibinfo {title} {Shaken and stirred: Random organization
  reduces viscosity and dissipation in granular suspensions}},\ }\href
  {https://advances.sciencemag.org/content/4/3/eaar3296} {\bibfield  {journal}
  {\bibinfo  {journal} {Science Advances}\ }\textbf {\bibinfo {volume} {4}}
  (\bibinfo {year} {2018})}\BibitemShut {NoStop}%
\bibitem [{\citenamefont {Sehgal}\ \emph {et~al.}(2019)\citenamefont {Sehgal},
  \citenamefont {Ramaswamy}, \citenamefont {Cohen},\ and\ \citenamefont
  {Kirby}}]{Sehgal2019_prl}%
  \BibitemOpen
  \bibfield  {author} {\bibinfo {author} {\bibfnamefont {P.}~\bibnamefont
  {Sehgal}}, \bibinfo {author} {\bibfnamefont {M.}~\bibnamefont {Ramaswamy}},
  \bibinfo {author} {\bibfnamefont {I.}~\bibnamefont {Cohen}},\ and\ \bibinfo
  {author} {\bibfnamefont {B.~J.}\ \bibnamefont {Kirby}},\ }\bibfield  {title}
  {\emph {\bibinfo {title} {Using acoustic perturbations to dynamically tune
  shear thickening in colloidal suspensions}},\ }\href
  {https://doi.org/10.1103/PhysRevLett.123.128001} {\bibfield  {journal}
  {\bibinfo  {journal} {Phys. Rev. Lett.}\ }\textbf {\bibinfo {volume} {123}},\
  \bibinfo {pages} {128001} (\bibinfo {year} {2019})}\BibitemShut {NoStop}%
\bibitem [{\citenamefont {Zhao}(2020)}]{zhao2020_phd}%
  \BibitemOpen
  \bibfield  {author} {\bibinfo {author} {\bibfnamefont {Y.}~\bibnamefont
  {Zhao}},\ }\emph {\bibinfo {title} {An Experimental Study of the Jamming
  Phase Diagram for Two-dimensional Granular Materials}},\ \href@noop {} {Ph.D.
  thesis},\ \bibinfo  {school} {Duke University} (\bibinfo {year}
  {2020})\BibitemShut {NoStop}%
\bibitem [{\citenamefont {Norden}(1973)}]{norden1973_report}%
  \BibitemOpen
  \bibfield  {author} {\bibinfo {author} {\bibfnamefont {B.~N.}\ \bibnamefont
  {Norden}},\ }\href@noop {} {\emph {\bibinfo {title} {On the compression of a
  cylinder in contact with a plane surface}}}\ (\bibinfo {year}
  {1973})\BibitemShut {NoStop}%
\end{thebibliography}%

\end{document}